\documentclass[trackchanges,twocolumn]{aastex7}

\usepackage[utf8]{inputenc}
\usepackage{subcaption}
\newcommand{\Ms}{M$_{\odot}$}

\newcommand{\Ni}{$^{56}$Ni}

\newcommand{\Co}{$^{56}$Co}

\newcommand{\mic}{$\mu$m}
\newcommand{\about}{$\sim$}
\begin{document}

%\title{Stochasticity in Stellar Yields Reflected in Theoretical Dust Masses Estimates Across all Type II Supernova Progenitors}

\title{Stochasticity in Stellar Yields Reflected in Supernova Dust Masses Across All Massive-Star Progenitors}

\author[orcid=0009-0006-2700-779X,sname='Purushothaman']{Archana Purushothaman}
\affiliation{Indian Institute of Astrophysics, 
100 Feet Rd, Koramangala, Bengaluru, Karnataka 560034, India}
\email{p.archana.ktr@outlook.com}

\correspondingauthor{Arkaprabha Sarangi} 
\author[0000-0002-9820-679X]{Arkaprabha Sarangi}
\affiliation{Indian Institute of Astrophysics, 
100 Feet Rd, Koramangala, Bengaluru, Karnataka 560034, India}
\email[show]{arkaprabha.sarangi@iiap.res.in}

\author[0000-0002-6517-7419]{S.~K.~Jeena}
\affiliation{Indian Institute of Astrophysics, 
100 Feet Rd, Koramangala, Bengaluru, Karnataka 560034, India}
\email{jeena.sk@iiap.res.in}  

%% Use the \collaboration command to identify collaborations. This command
%% takes an optional argument that is either a number or the word "all"
%% which tells the compiler how many of the authors above the command to
%% show. For example "\collaboration[all]{(DELVE Collaboration)}" wil include
%% all the authors above this command.
%%
%% Mark off the abstract in the ``abstract'' environment. 
\begin{abstract}
Massive stars, ending their lives as supernovae (SNe), are among the primary sources of dust in galaxies. In this study, we derive theoretical upper limits on dust masses as a function of SN progenitors, assuming non-rotating single stars of solar metallicity, with initial masses between 9 and 120~\Ms. Based on previously established models of dust formation chemistry in core-collapse SNe (CCSNe), we find that O-rich dust, particularly silicates and silica, dominates the dust budget, with masses ranging from 0.02 to 1.43~\Ms, and that the total mass of O-rich dust increases with progenitor mass. C-rich amorphous carbon and silicon carbide dust are significant for lower-mass progenitors (10--15~\Ms), but their mass never exceed 0.05~\Ms. For progenitors up to 30~\Ms, we provide best-fit functions describing the masses of O-rich dust, C-rich dust, and CO molecules. A large stochastic variation is found in the predicted masses of silicate dust, which correlates with the randomness of shell-merger events in the pre-explosion phases of massive stars. Furthermore, we show that the dust mass for a given progenitor can vary by a factor of 2--5, reflecting differences in pre-explosion abundance distributions predicted by the different stellar evolution models. We emphasize that the final dust yield in SNe is primarily determined by stochastic stellar yields and uncertainties in pre-explosion nucleosynthesis, while explosion properties mainly influence the timescales of dust formation.

\end{abstract}

%% Keywords should appear after the \end{abstract} command. 
%% The AAS Journals now uses Unified Astronomy Thesaurus (UAT) concepts:
%% https://astrothesaurus.org
%% You will be asked to selected these concepts during the submission process
%% but this old "keyword" functionality is maintained in case authors want
%% to include these concepts in their preprints.
%%
%% You can use the \uat command to link your UAT concepts back its source.

%% From the front matter, we move on to the body of the paper.
%% Sections are demarcated by \section and \subsection, respectively.
%% Observe the use of the LaTeX \label
%% command after the \subsection to give a symbolic KEY to the
%% subsection for cross-referencing in a \ref command.
%% You can use LaTeX's \ref and \label commands to keep track of
%% cross-references to sections, equations, tables, and figures.
%% That way, if you change the order of any elements, LaTeX will
%% automatically renumber them.

\section{Background} \label{sec:Intro}

In the last few decades, SNe, especially  CCSNe, have been confirmed as sources of dust in galaxies \citep{Matsuura2015, szalai_2019, wesson_2021, maria_2021, subrayan_2026}. Mid-infrared (mid-IR) observations using the James Webb Space Telescope (\textit{JWST}) over the last two years have revealed dust masses exceeding 0.02 \Ms\ in several decade-old SNe \citep{Shahbandeh2023, shahbandeh_2024, sarangi_2025a}. SNe such as SN~2004et and SN~2005af were found to be dominated by C-rich, amorphous carbon dust \citep{shahbandeh_2023, sarangi_2025a}. On the other hand, SN~2017eaw and SN~2005ip showed signs of O-rich, silicate dust features \citep{shahbandeh_2023, shahbandeh_2024}. The location of dust around the SN remains ambiguous, with possible sites including dust in the surrounding CSM and/or newly formed dust in the ejecta or post-shock cooling gas \citep{shahbandeh_2024}.

In general, theoretical models of SN dust formation predict rapid dust synthesis to grow dust masses to \about 0.1~\Ms\ within a couple of years \citep{sarangi2018book, sluder2018}. They do not converge well with the observed, gradual, and controlled rates of dust formation in SN ejecta that continue for several decades \citep{wesson_2021, gal14}. If dust is produced early and/or in optically thick clumps, IR fluxes may not accurately reflect the dust masses present in the system \citep{dwek_2019}. CCSNe subclasses such as Type II-P, IIn, IIb, Ib, and Ic are expected to form dust at different rates or in different locations; however, observations alone are insufficient to distinguish between these scenarios. The final mass of dust is also intricately related to the late-stage evolution of massive stars -- how much mass the stars lose prior to their explosions, if dust will form before the explosion in the wind, and if it will survive after the explosion.   

In the early universe, SNe from the short-lived massive stars are expected to be the primary source of dust \citep{Schneider2023, dwe11}. They are also known to be major dust producers in local galaxies \citep{Matsuura2011}. In this study, we derive the theoretical upper limits of dust masses and their chemical compositions across thinly spaced bins of progenitor masses from 9 to 120 \Ms\ of solar metallicity. The results are valid irrespective of the SN subclass and explosion properties (eg, explosion energy, \Ni\ mass, clumpiness).

\begin{figure*}
    \centering

    \begin{subfigure}{\textwidth}
        \includegraphics[width=\linewidth]{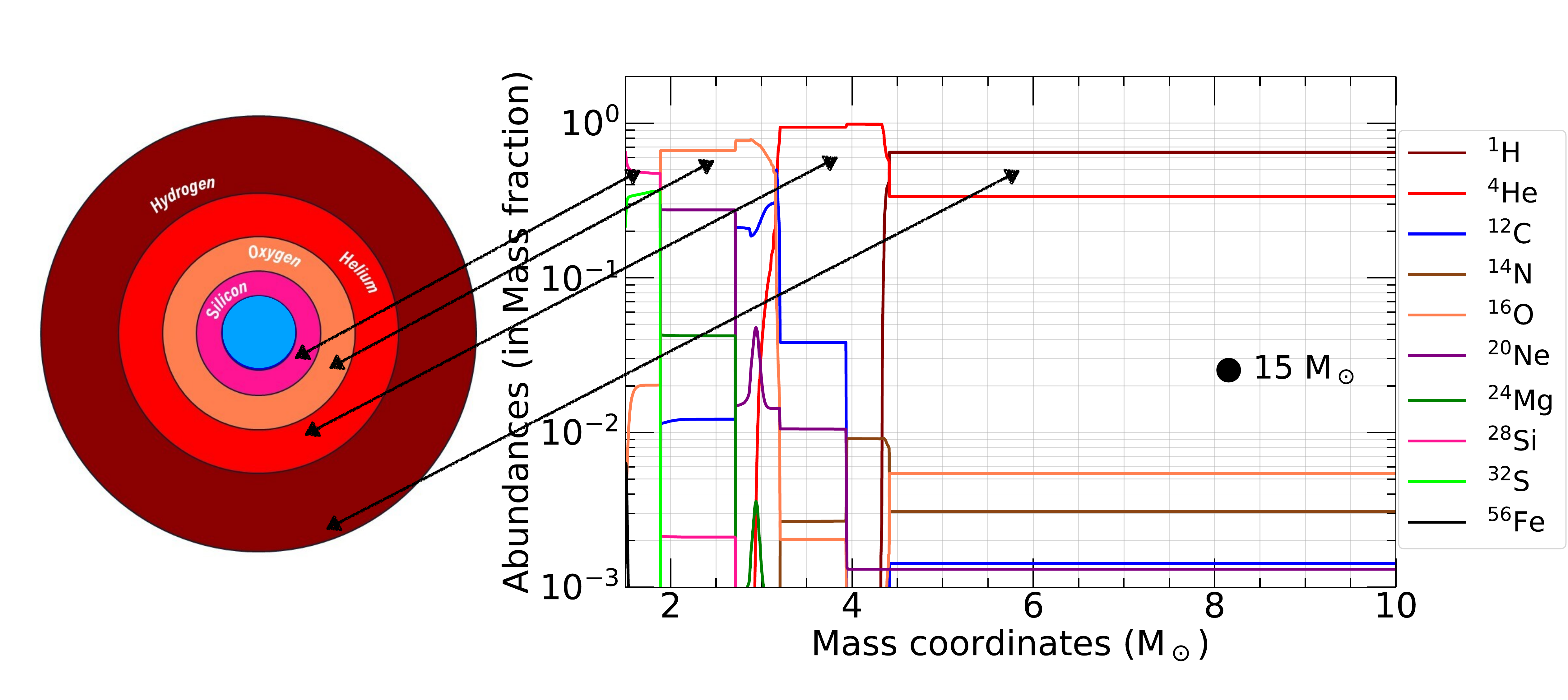}
    \end{subfigure}

     \begin{subfigure}{0.48\textwidth}
        \includegraphics[width=\linewidth]{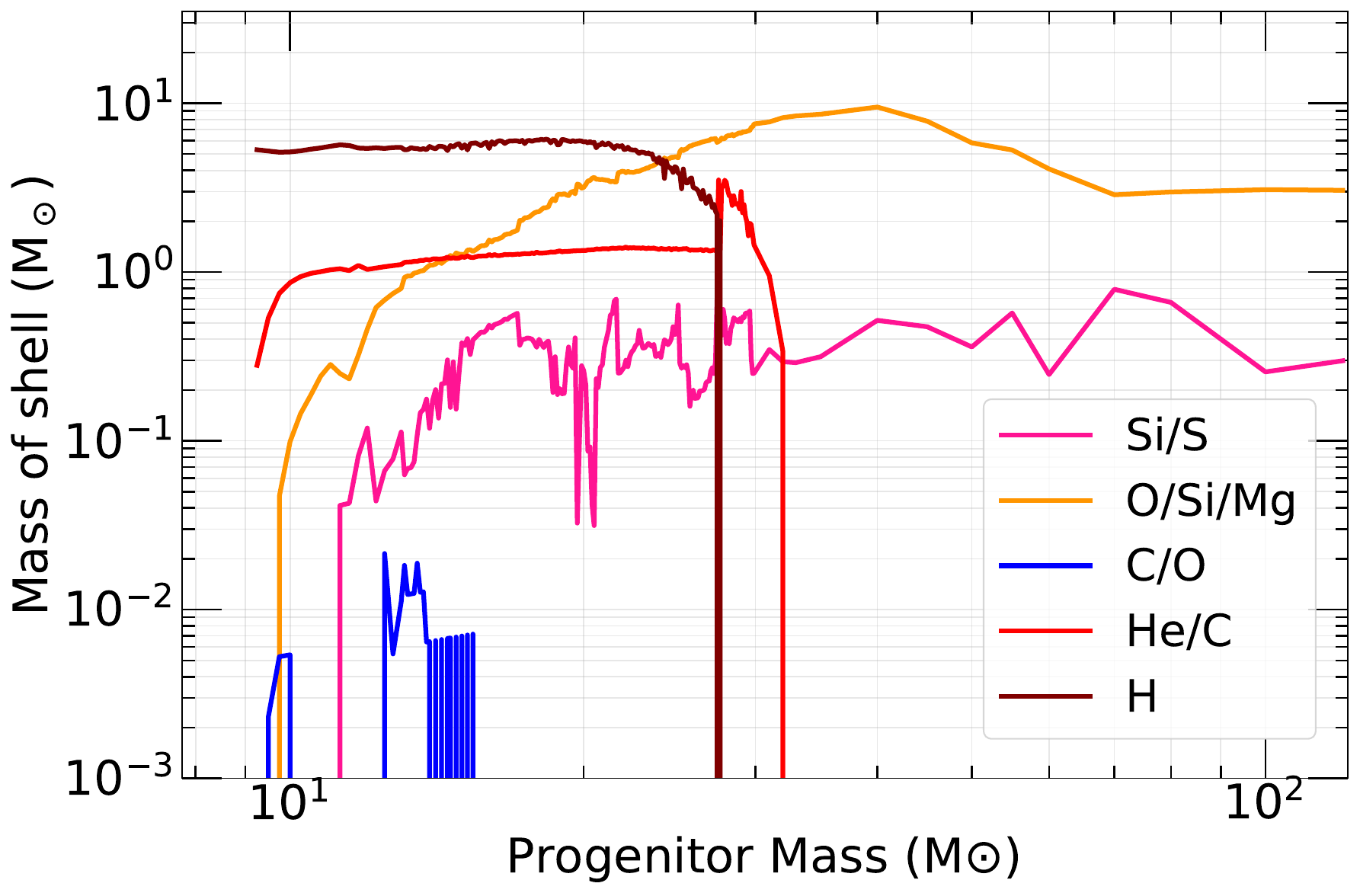}
    \end{subfigure}
    \hfill
    \begin{subfigure}{0.48\textwidth}
        \includegraphics[width=\linewidth]{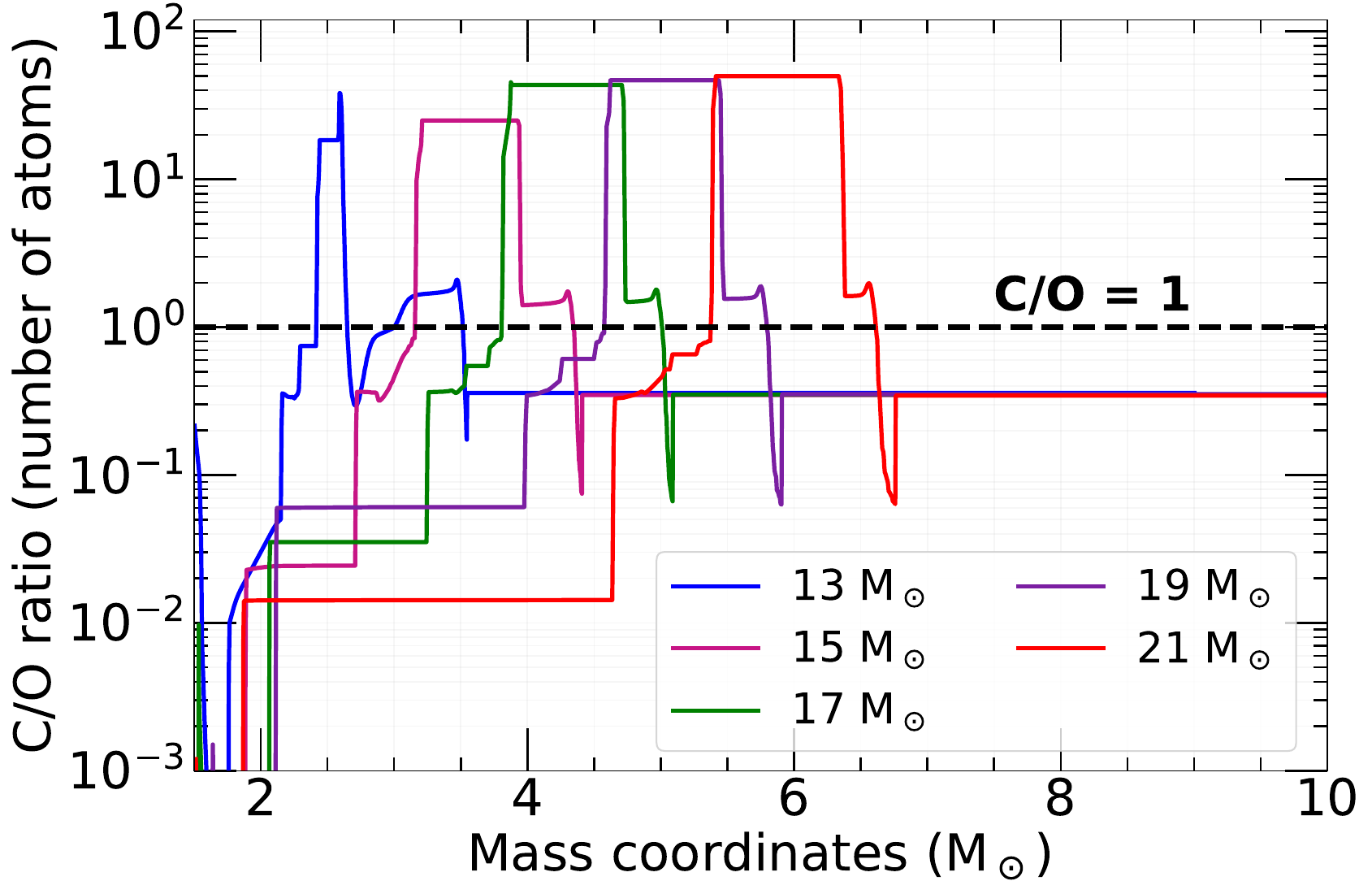}
    \end{subfigure}

    \begin{subfigure}{0.48\textwidth}
        \includegraphics[width=\linewidth]{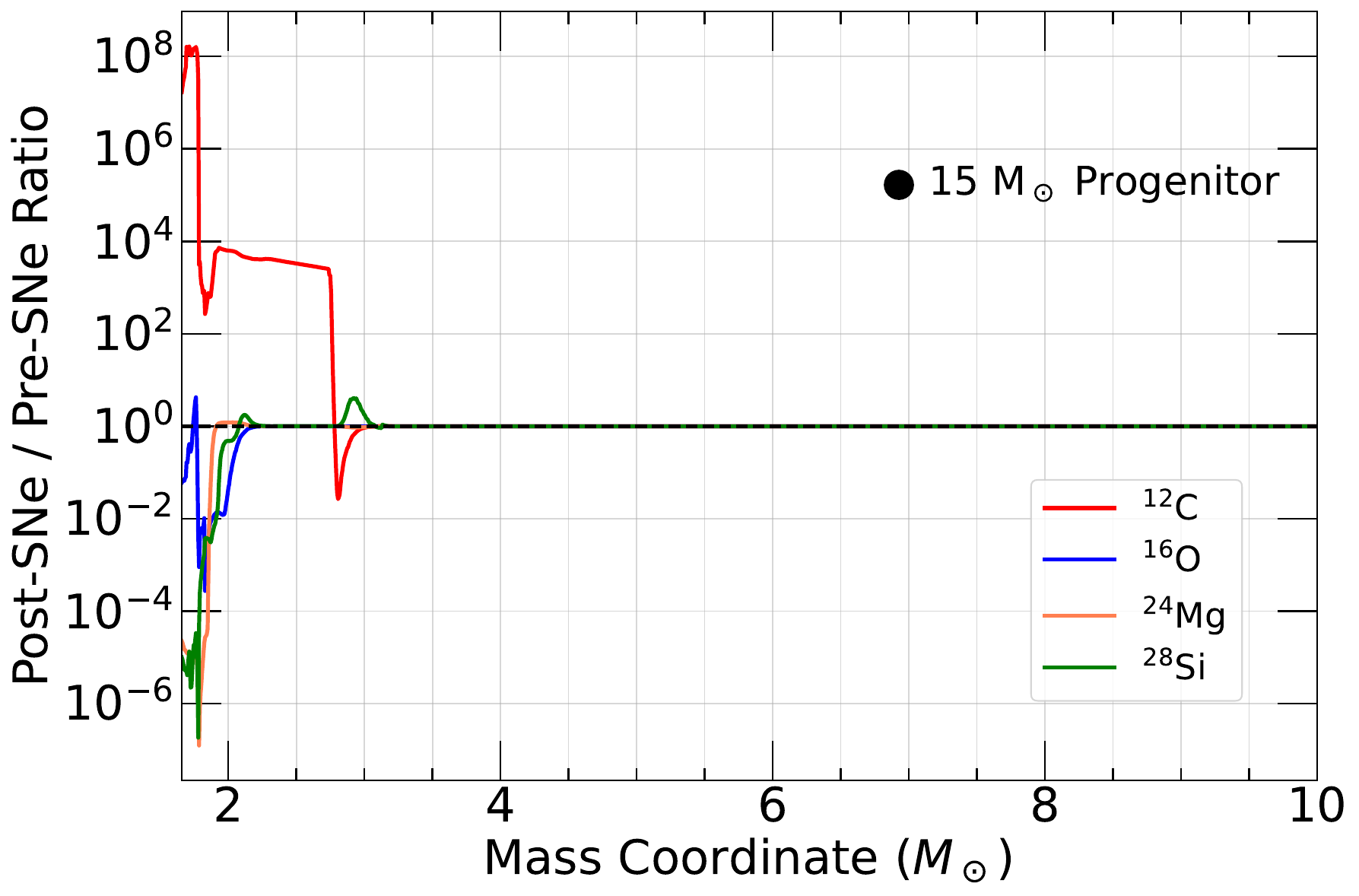}
    \end{subfigure}
    \hfill
    \begin{subfigure}{0.48\textwidth}
        \includegraphics[width=\linewidth]{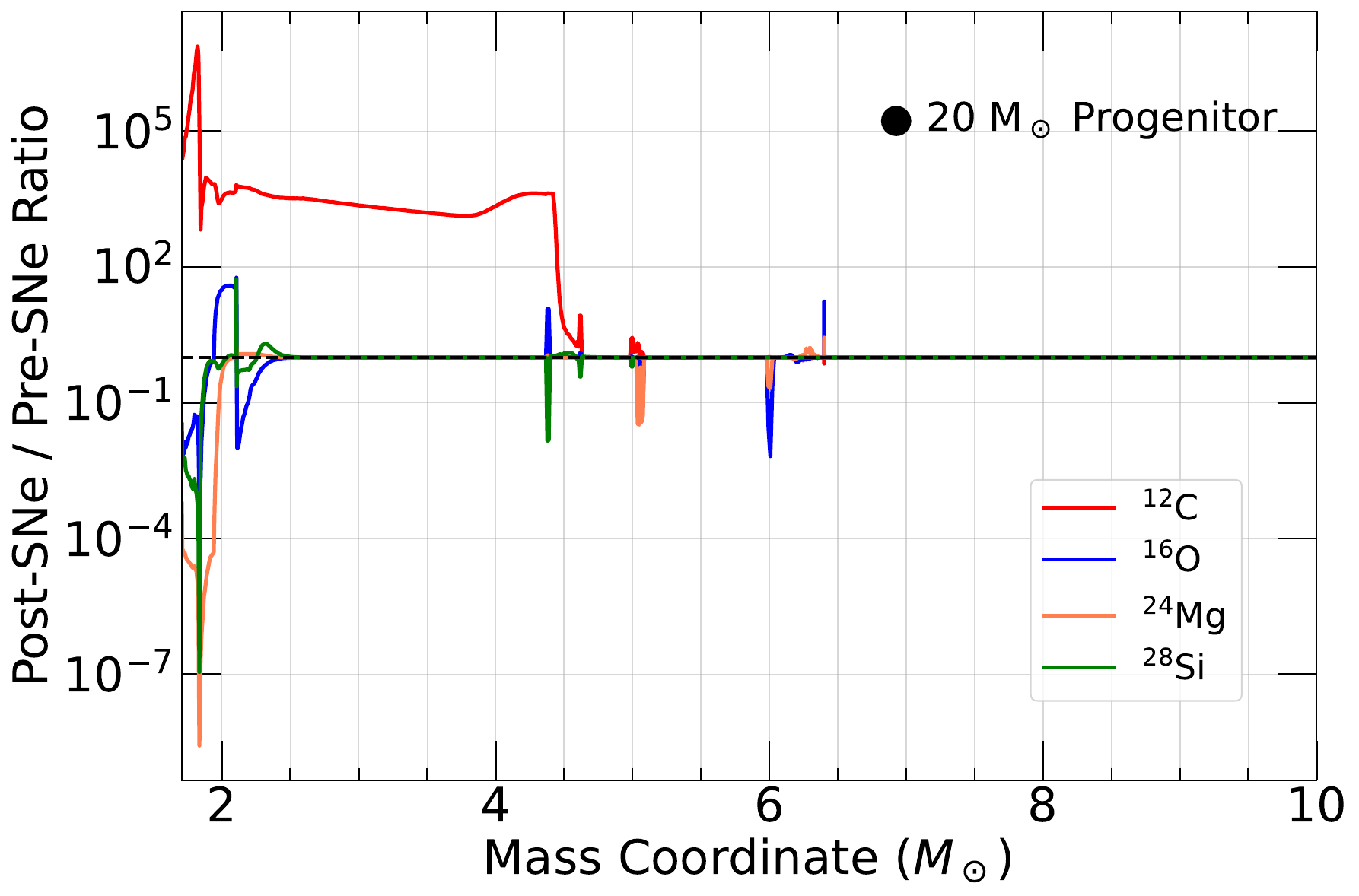}
    \end{subfigure}

\caption{Top: Stratified zones of a 15 $M_\odot$ progenitor from \citet{Sukhbold2016}, illustrating the classic ``onion-shell" layering structure. The detailed mass fraction distribution of key elements within each zone is also illustrated, highlighting compositional transitions between zones. \quad Middle left: Masses of zones across all progenitors from \citet{sukhbold_2016}. The He/C and the H zones become negligible/non-existent around 32 $M_\odot$ progenitor onward, and the O/Si/Mg zone shows a steady increase across all progenitors. \quad Middle right: Carbon to Oxygen (C/O) ratio by number of atoms for various \Ms\ progenitors. The dotted black line represents $C/O =1$ . The zones where $C/O > 1$ will produce predominantly C-rich dust and  $C/O < 1$ produce O-rich dust. \quad  Bottom: Ratio of post-explosion to pre-explosion elemental abundances using stellar evolution model \texttt{KEPLER} \citep{Woosley_1995, Woosley2007} for a typical 15 $M_\odot$ (left) and a 20 $M_\odot$ (right) initial mass progenitor, respectively.} 
    \label{fig:zones_KEPLER}
\end{figure*}

\section{Rationale of this study}

The objective of this study is to derive an empirical relation between dust mass, dust composition and progenitor mass. Detailed chemical modeling of dust formation in SNe \citep{sar15, sarangi2018book} tracks the formation of molecules, clusters, and condensation of large clusters into dust. We use those results to constrain the abundances and chronology of molecules and dust synthesis in SN ejecta. This enables us to determine the upper limit of each dust type for a given progenitor based on yields obtained from stellar evolution models \citep{woo95, paxton_2011}.  

The paper is arranged in the following order: In Section \ref{sec:SNe_progenitors}, we discuss the observational and theoretical perspectives of SN progenitors. Section \ref{sec:dust_conditions} describes the conditions for dust formation after those progenitors explode as SNe. In Section \ref{sec_upper_limit} we quantify the upper limits of dust masses derived from this study. Section \ref{sec_compactness} focuses on the stochastic nature of derived dust masses, with reference to stellar evolution processes, namely shell merging and compactness of the stellar models. Thereafter, in Section \ref{sec_fit_dustmass}, we provide analytic functions that describe dust masses as a function of progenitor masses. In light of our results, in Section \ref{sec_MESA}, we show the sensitivity of our study in comparison to other stellar evolution models. Finally, Section \ref{sec_discussion} discusses the findings and their relevance to SN observations.

\section{Supernovae Progenitors} \label{sec:SNe_progenitors}
Massive stars, with initial masses larger than 8 \Ms\ are considered to be the progenitors of CCSNe \citep{Heger_2003,2003A&A...404..975M}. There are large uncertainties in determining the progenitor masses of SNe from pre-SN imaging or light curve modeling. As an example, for the nearby SN~2023ixf, the reported progenitor is found to be of masses spanning a wide range between 9 to 22~\Ms\ \citep{niu_2023, liu_2023, pledger_2023, kilpatrick_2023, jencson_2023, neustadt_2024, xiang_2024, vandyk_2024, ransome_2024, qin_2024, soraisam_2023, hsu_2024, bersten_2024, moriya_2024, singh_2024, ferrari_2024}. It is safe to say, we have no CCSNe with an accurate determination of their progenitor mass.

Given their unique evolutionary channels, each progenitor will differ in its abundance distribution, final yields, and explosion properties. Each of these factors are significant for dust formation. Stellar evolution codes like \texttt{KEPLER} \citep{woo02, rau02, Woosley_1995, Woosley2007, weaver_1978} and \texttt{MESA} \citep{paxton_2011, paxton_2013, paxton_2015} simulate the evolution of a progenitor star from the pre-main sequence, throughout all the burning phases and nucleosynthesis channels till the point of CC.

The elemental yields in the stellar evolution models show considerable variations based on progenitor masses, especially in their post-main sequence phases. We probe a large sample of non-rotating single stars of solar metallicity, as progenitors of CCSNe. For that purpose, we use pre-explosion yields from a finely spaced grid of 200 progenitors with main sequence masses between 9 to 120 \Ms\ using the simulations of \cite{sukhbold_2016} \footnote{KEPLER data: \url{https://wwwmpa.mpa-garching.mpg.de/ccsnarchive/data/SEWBJ_2015/index.html}}. These models were obtained using the stellar evolution code \texttt{KEPLER}, and in the following text, where we have cited this code, we specifically refer to the models of \cite{sukhbold_2016}. 

CCSNe progenitors are expected to be less than 25 \Ms\ \citep{smartt_2009}, and in that range (9-25 \Ms) the grid has a spacing of 0.1--0.5 \Ms. Since our study focuses on the theoretical upper limit of dust masses, we also present some larger progenitors using the available data. In Figure \ref{fig:zones_KEPLER} (top), we show the stratified internal chemical profile of a 15 \Ms\ star from this data set. In the middle panel, we show the masses of the individual zones at the time of core-collapse, as a function of the initial mass of the star. The ratio of C to O atoms in the ejecta is shown, since it is a key composition determining dust formation sequence, explained in more detail in Section \ref{sec:dust_conditions}. 

Only pre-explosion models were available for this large sample of progenitor masses. We consider that for the alpha elements, which are the primary constituents of dust, the pre- and post-explosion abundance profiles are comparable. For confirmation, we simulated a 15 and a 20 \Ms\ initial mass star until core-collapse using the same initial conditions as \cite{sukhbold_2016}, and the same stellar evolution codes \texttt{KEPLER}, as we show in Figure \ref{fig:zones_KEPLER} (bottom).

\section{Conditions for dust formation} \label{sec:dust_conditions}

Dust formation models suggest that SN ejecta undergo nucleation and condensation phases, with efficient dust formation occurring within the first few years post-explosion \citep{cherchneff2009,sarangi2018book, sarangi_2022b}. The process involves gas-phase chemistry in the nucleation phase, leading to the synthesis of molecules and gas-phase clusters. This is followed by condensation, where these precursor molecules grow into larger dust grains and clusters through coagulation, coalescence, and accretion on the surface. 

The chemical abundance in the ejecta has a significant influence on the nature of the dust formed. 
In a similar approach to our previous dust models \citep{sar13, sar15, sarangi_2022b}, we assume that the ejecta is characterized by stratified mass zones (Si/S, O/Si/Mg, He/C, H) with initial chemical composition derived using stellar evolution models \citep{sukhbold_2016, sukhbold_2018}. The gas within each zone/layer is microscopically mixed, and there is no macroscopic mixing between the zones. Based on the unique chemistry of each zone (due to variations in abundances, temperature profile, densities, clumpiness, radioactive decay energy etc.), molecules and dust grains differ in timescale of formation, final mass, chemical composition, and size distributions. As previous SN dust chemistry models \citep{sarangi_2022b}, we have considered two major dust categories, O-rich dust -- silicates, alumina, and C-rich dust -- amorphous carbon and silicon carbide.

% The various dust particles have distinct properties such as their sources, temperature profile, survivability, dust grain size distribution and structure, IR emission signatures, etc.  The major molecular precursors to dust grains that typically form in the aftermath of a supernova can be classified into:

Oxygen-rich dust is known to form in regions where silicon and oxygen abundances are high. Mg-silicates, mainly of chemical type [Mg$_2$SiO$_4$]$_n$ are expected to form in SN ejecta. All the zones are efficient in forming silicate dust. However, given the abundances, the majority of silicates are formed in the O/Si/Mg zone. Silicates are often known to be the most abundant dust in SNe, as detected in SN~2005ip, SN~2017eaw or Cas A \citep{shahbandeh_2023,shahbandeh_2024, arendt_2014} by their signature $9.7$ and $18$ \mic\ features \citep{breemen_2011}. Theoretical models have predicted silicates to start forming rapidly in the ejecta as soon as a year post-explosion \citep{sarangi_2025a, sarangi2018book}.

\cite{cherchneff_2025} proposes a nucleation channel of silicate dust that includes the formation of silica (SiO$_2$), and MgO molecules. There is a large uncertainty on the likelihood of MgO molecule formation, given that it requires crossing large energy barriers, whereas the dissociation of MgO molecules is relatively much more efficient \citep{bhatt_2007, bell_2025}. 

%Among the SNe detected by \textit{JWST}, such as SN~2017eaw, SN~2004et, SN~2005af, SN~1980K, SN~2005ip, SN~1993J, or SN~1995N \citep{shahbandeh_2023, shahbandeh_SN2005ip, sarangi_2025a, szalai_2025, szanna_2024, clayton_2025}, many show features of Mg-silicates, while silica or Mg-depleted silicates were not reported. On the other hand, in SN remnant Cas~A and G54.1+0.3, a 21 \mic\ feature is attributed to silica dust \citep{rho_2008, rho_2018}. It remains unclear if [SiO$_2$]$_n$-type compounds may result from the dissociation of the silicates by the reverse shock.

The presence of a 21 \mic\ feautre in SN remnant Cas~A and G54.1+0.3 was attributed to Mg-depleted silicates such as silica (SiO$_2$), and  Mg$_{0.7}$SiO$_{2.7}$ \citep{rho_2008, rho_2018, arendt_2014, temim_2017}. Such a feature, however, is not present in the SNe detected by \textit{JWST}, such as SN~2017eaw, SN~2004et, SN~2005af, SN~1980K, SN~2005ip, SN~1993J, or SN~1995N \citep{shahbandeh_2023, shahbandeh_SN2005ip, sarangi_2025a, szalai_2025, szanna_2024, clayton_2025}. The large optical depths associated with dust in the ejecta could be one reason behind the suppression of any spectral feature, as postulated by \cite{dwe15, sarangi_2022b}. Our study is based on dust synthesis chemistry of \cite{sar13, sar15, sarangi_2022b}, which does not predict efficient synthesis of silica (SiO$_2$) dust. However, in this paper, our focus is on the upper limits of dust masses, so we include the contribution of Mg-depleted silicates as well to the total dust mass. 

\begin{figure*}
    \centering
        \includegraphics[width=\linewidth]{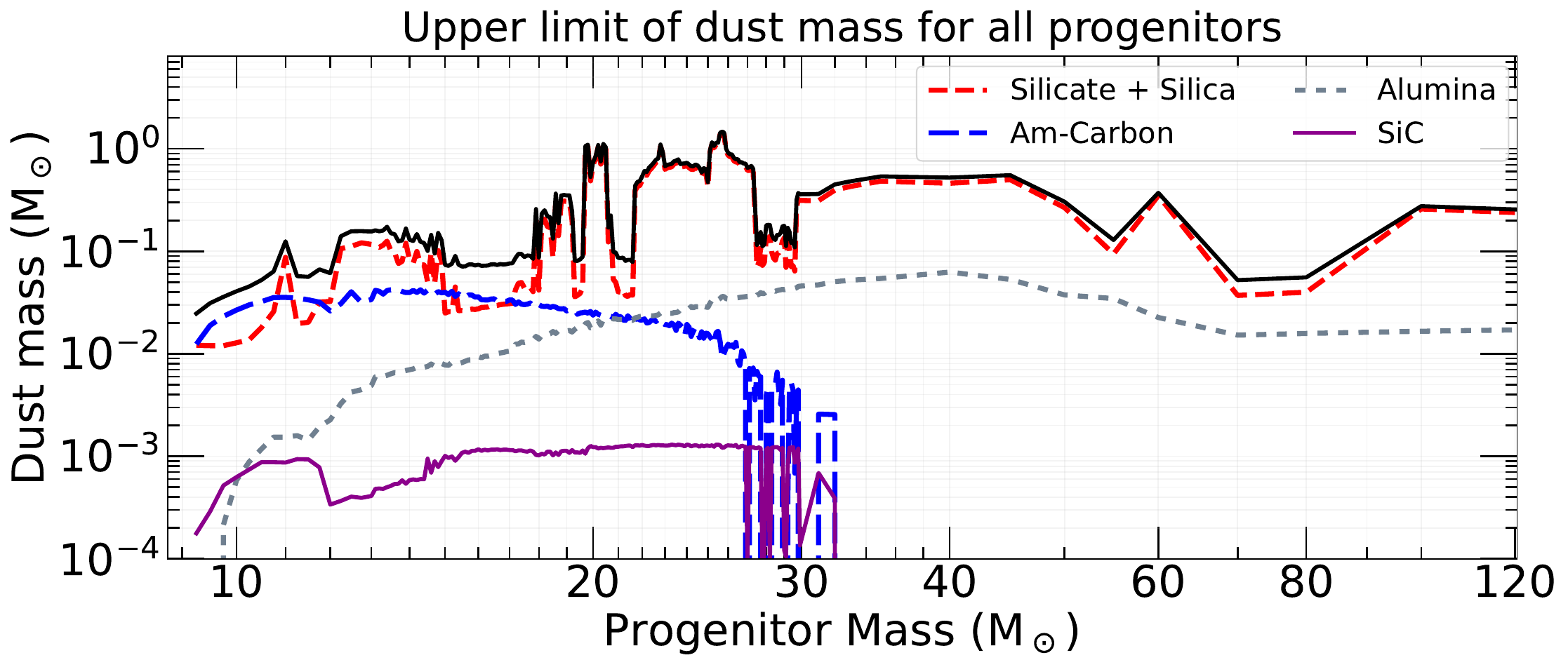}

    \caption{Upper limits of silicates, alumina, amorphous carbon and silicon carbide dust across all progenitors are shown. The C-rich dust are represented by amorphous carbon (Am-Carbon in blue) and silicon carbide (in purple). The O-rich dust is the combined mass of the silicates + silica (red), along with alumina (grey) dust, which are the most abundant species of the O-rich dust. The upper limit of the total dust mass, which is the sum of the C-rich and O-rich dust masses, is plotted in black.}
    \label{fig:Dust_KEPLER}
\end{figure*}

\begin{figure*}
    \centering
        \includegraphics[width=\linewidth]{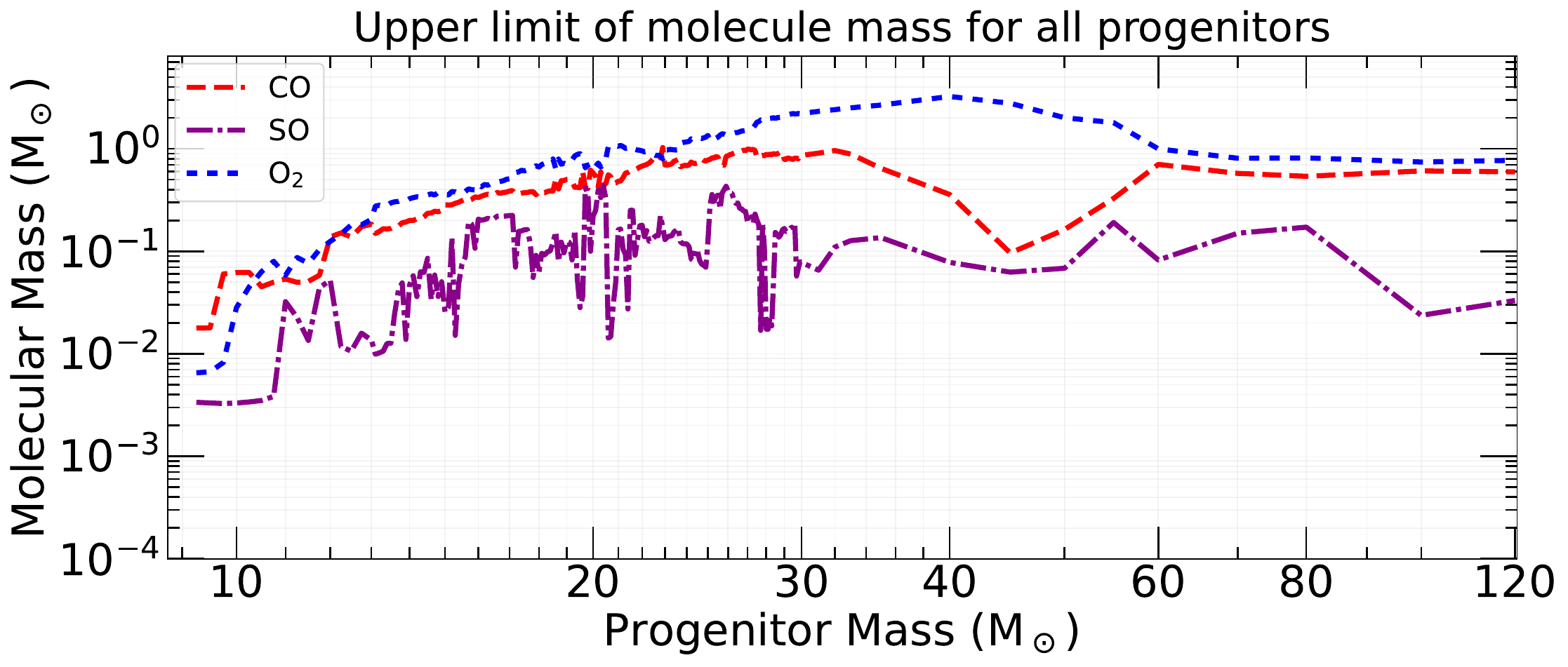}
        \caption{Upper limits of CO, SO, and $O_2$ molecules across all projenitors. The CO, SO and $O_2$ molecules are plotted by red, purple and blue colors respectively. }
    \label{fig:molecules_KEPLER}
\end{figure*}

Apart from silicates, aluminum oxide or amorphous alumina, of chemical type [Al$_2$O$_3$]$_n$, is also an abundantly formed O-rich dust species. Importantly, though, the abundance of Al is altered significantly during the explosive nuclear burning phase, and hence we cannot use the pre-explosion abundance profile for Al. We have used the O to Al ratio in the O/Si/Mg zone (where alumina forms) from the post-explosion profile of a $15$~\Ms\ progenitor, modeled using the \texttt{KEPLER} code, and scaled it for other progenitors as well. This approximation may induce errors, but since alumina is not the most abundant dust type, we expect the error to be insignificant. 

\begin{figure}
    \centering
        \includegraphics[width=3in]{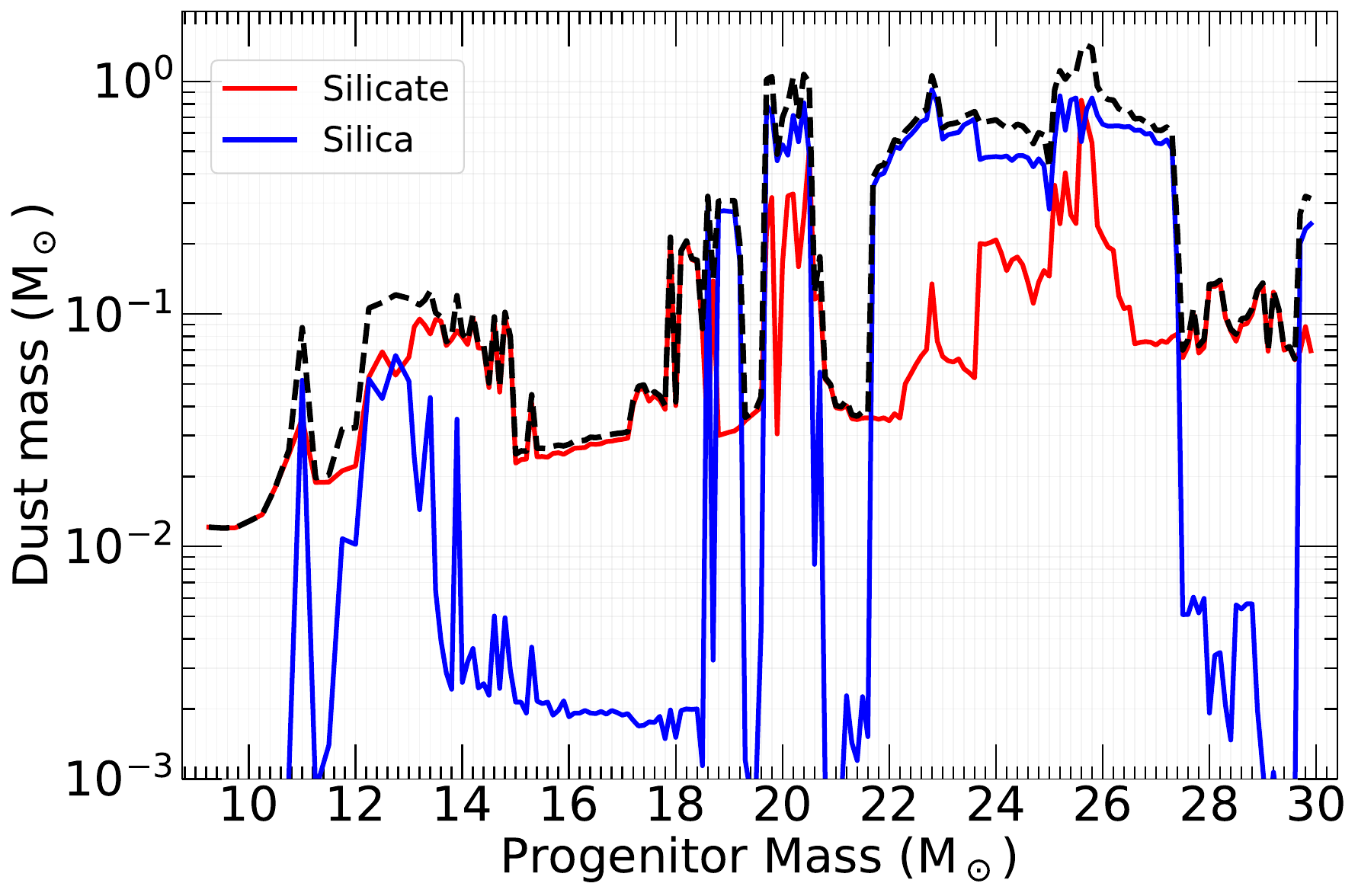}
    
    \caption{The masses of silicates (chemical type [Mg$_2$SiO$_4$]$_n$) are compared to a possible upper-limit of silica (chemical type [SiO$_2$]$_n$), across all progenitor masses. The sum of the two Si-rich dust is shown by the black dashed line.}
    \label{fig:Dust_silica}
\end{figure}

Amorphous carbon and silicon carbide are the C-rich dust component in SNe. A critical factor to form C-rich dust is the carbon-to-oxygen ratio of the ambient/local medium.  In the zones, where $\rm{C/O}<=1$, all the C-atoms are found to be locked in CO molecules. CO molecules are efficiently formed in the ejecta and do not break down easily due to their large binding energy. However, that leaves no scope to synthesize C-dust in such zones, which are formed through clusters of pure carbon chains and rings. In massive stars, the He-rich region, as shown in Fig. \ref{fig:zones_KEPLER} (top, bottom-right), is uniquely C-rich, and allows to form amorphous carbon dust when suitable conditions are met \citep{sar13}.

Progenitors of various masses evolve differently, leading to a variation in the sizes of the shells, as we show in Fig. \ref{fig:zones_KEPLER} (bottom-left). The mass of the shells directly impacts the composition of dust, since O-rich dust is formed in the inner Si/S and O/Si/Mg zones, while C-rich dust in formed in the He/C zone. The mass of the He/C zone remains almost identical over all progenitor masses until $25$~\Ms, however, the C/O ratio does show significant variation (see C/O ratio in Figure \ref{fig:zones_KEPLER}, middle-right). The O/Si/Mg zone, otherwise also called the O-core, increases in mass with progenitor mass, thereby making O-rich dust grains more likely to be abundant. 

Extensive studies of dust formation chemistry reveal that SiO and CO molecules are abundantly formed in the SN ejecta \citep{sar13, sar15, sarangi2018book}. SiO acts as a precursor or initiator to dust formation, leading to the pathway of forming silicate dust. The CO molecule, on the other hand, is an adversary or competitor to dust production, as it sequesters the carbon and oxygen atoms, which are key constituents of dust. In addition, O$_2$ and SO molecules also form in significant quantity in the metal-rich ejecta.

\begin{figure}[t] % single float environment
    \centering
    \includegraphics[width=24em]{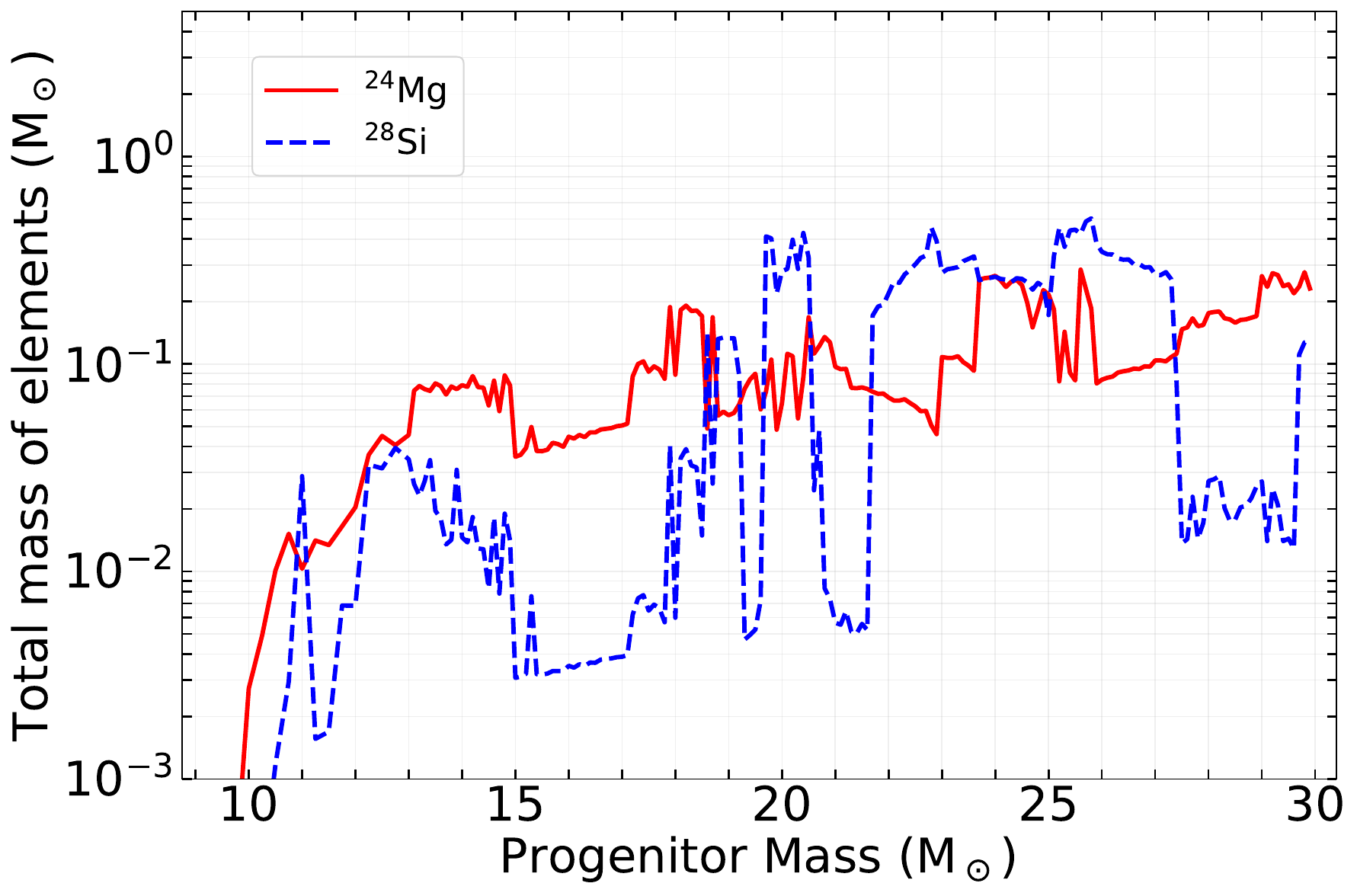}
    \includegraphics[width=\columnwidth]{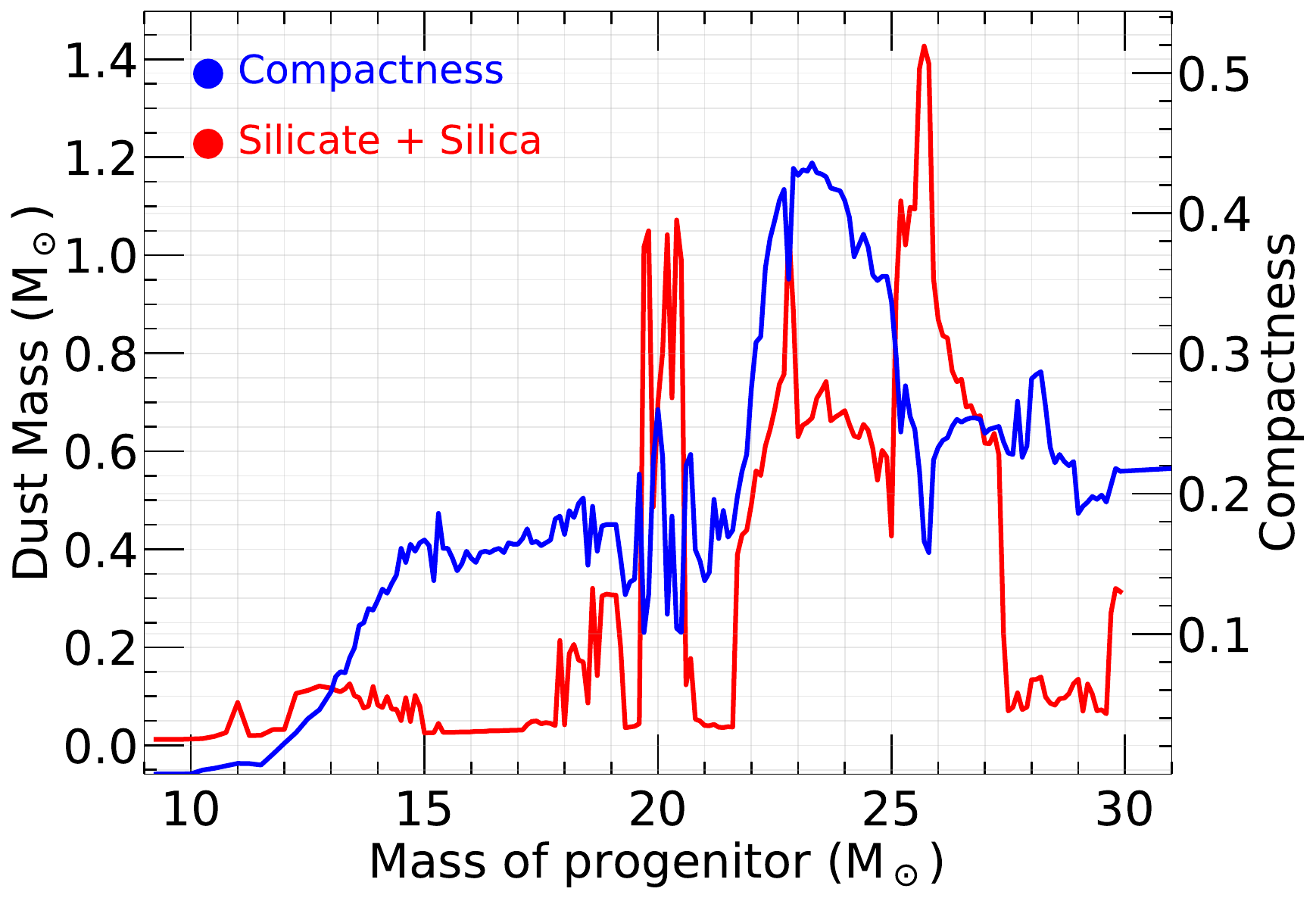}
    \includegraphics[width=\columnwidth]{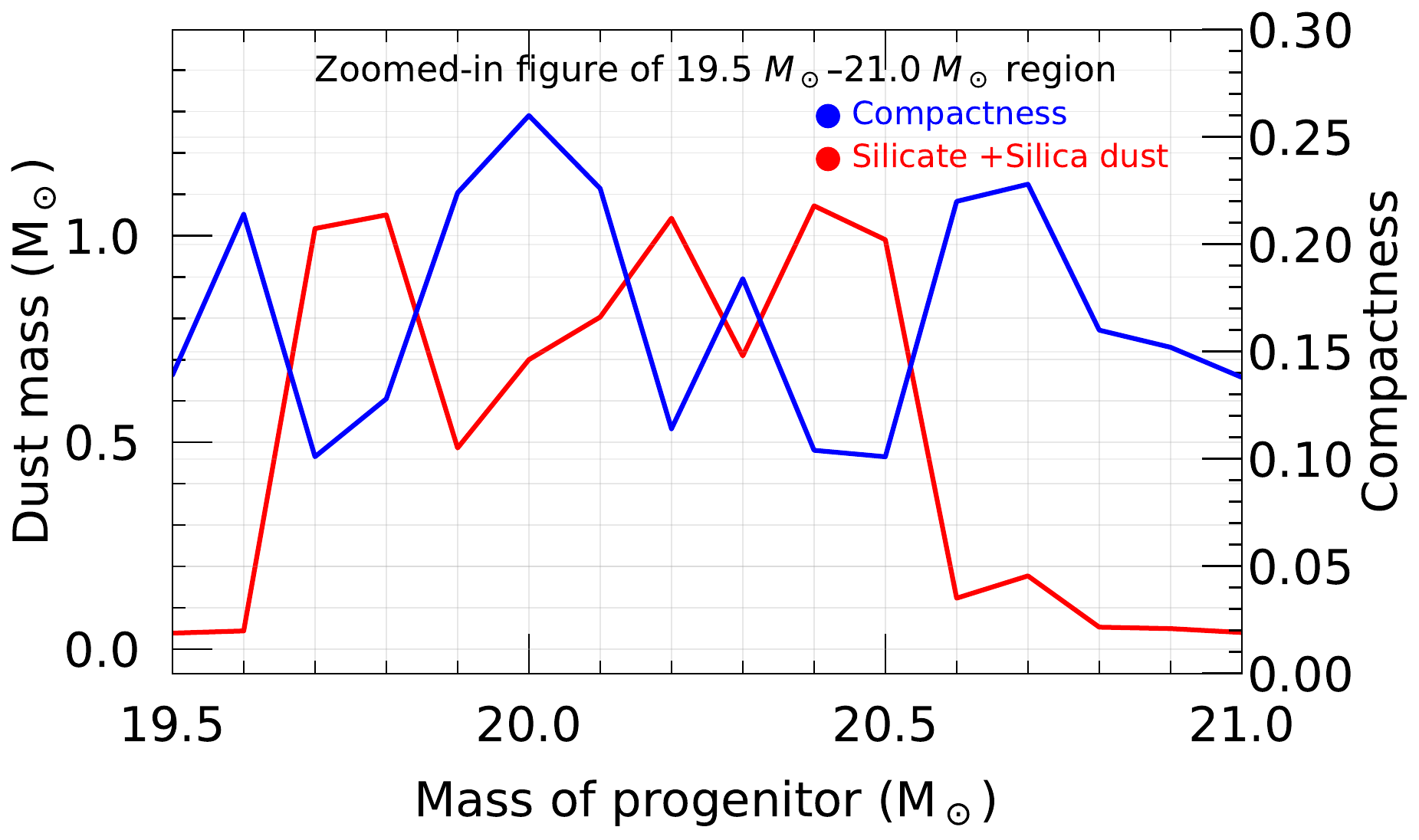}
    \caption{Top: Abundances of Si and Mg atoms in the O/Si/Mg zone across all progenitors up to 30 \Ms\ from \cite{Sukhbold2016}. See Section \ref{sec_compactness} for context. \quad Middle: The correlation between compactness parameter (in blue) and mass of silicate + silica dust (in red). Compactness parameter (Equation \ref{eq:compactness}) is used as a tracer of the shell merging scenario explained in Section \ref{sec_compactness}. \quad Bottom: Zoomed-in view of the 19.5 to 21 \Ms\ region, showing compactness parameter and mass of silicate + silica dust}. 
    \label{fig:KEPLER_elemental_mass_compactness}
\end{figure}

\section{Upper limit of dust masses} \label{sec_upper_limit}

The epoch when dust formation commences, and the rate at which it proceeds, is influenced by the local conditions such as density, clumpiness, temperature etc. However, irrespective of the rate of dust formation, previous models have found for the cases of 12, 15, 19 or 20 \Ms\ progenitors \citep{sar13, sar15, sarangi2018book, sarangi_2022b}, that the final dust mass in each zone is limited by the abundance of the least abundant element among the dust constituents in each zone, after CO molecules are formed. To explain with an example -- for silicate dust in the O/Si/Mg zone, first CO molecules are formed, which lock up C and O atoms in equal proportions. Following that, clusters of Mg-silicates are formed in mass that is limited by the least abundant of the Mg, Si, or O atoms, in the proportion of the atomic constituents of [Mg$_2$SiO$_4$]$_n$ (for integer $n>2$). If the Mg-atoms are consumed this way, before Si-atoms are exhausted, Mg-depleted silicates such as silica dust will form in the ejecta. In the same way, in the He/C zone, once all the CO molecules are formed, the remainder of the C atoms form chains and rings, ultimately condensing to amorphous carbon dust. The mass is limited by the remaining C-atom in the gas, post CO molecule formation. SiO molecules, formed in the gas as early as CO molecules, ultimately deplete in silicate dust. Other O-bearing atoms, such as O$_2$, AlO, MgO, SO, etc., may still form in the gas if there are residual O atoms. 

Based on this chronology, we can estimate the dust masses of each type that is formed in each zone, which will be the theoretical upper limit. 

Using the formalism described above, we find the upper limit of O-rich (silicates, silica and alumina) and C-rich (amorphous carbon, silicon carbide) dust for all progenitors between 9-120 \Ms\ (main sequence mass), which were simulated by \cite{sukhbold_2016} using the \texttt{KEPLER} code. Simultaneously, for the same models, the upper limit for the molecular masses are also derived. 

In Figure \ref{fig:Dust_KEPLER}, we show the masses of dust for each progenitor. A 9~\Ms\ progenitor produces 0.025~\Ms\ of dust, which is the smallest dust mass in this set, while a 25.5~\Ms\ star with a dust mass of 1.43 \Ms\ is the highest mass of dust we report. As expected, the mass of O-rich dust (silicates, silica, alumina) increases with progenitor mass, since the size of the O-core (O/Si/Mg zone) is found to increase proportionately (see Figure \ref{fig:zones_KEPLER}, middle-left). Mass of C-rich, amorphous carbon and silicon carbide dust, is found to vary between 0.012 to 0.043~\Ms, with the maximum occurring in progenitors with initial masses between 13 and 14~\Ms. The C-rich dust is mostly in the form of amorphous carbon; the mass of silicon carbide does not exceed much beyond 10$^{-3}$ \Ms. The mass of C-rich dust drops to negligible quantities for stars with initial mass larger than 26 \Ms. O-rich dust, namely silicates, dominates the dust composition almost always, for progenitors with initial mass above 12 \Ms, except for a few cases between 15 and 16 \Ms.  

The molecular masses are shown in Figure \ref{fig:molecules_KEPLER}. CO molecules, are formed consistently in all progenitors, where the mass (upper limit) increases gradually from 0.02 \Ms\ to as large as 1 \Ms\ for a progenitor with an initial mass of 26~\Ms. Due to the large abundances of O-atoms, we find the molecular masses of O$_2$ to be in the range between 0.01 and 3 \Ms. The SO molecule masses are between 10$^{-3}$ to 0.2 \Ms. We find a large fluctuations in the masses of SO molecules as a function of initial stellar masses, which reflects to the uncertainty in the abundances of S-atoms. In this study, we aim to find the maximum possible dust mass, and in that formalism we are left with almost no SiO molecules in the gas, since they efficiently condense to silicate dust. 

%CO molecules, shown in Figure \ref{fig:Dust_fit}(bottom-left), are formed consistently in all progenitors, where the mass (upper limit) increases gradually from 0.02 \Ms\ to as large as 1 \Ms\ for a progenitor with an initial mass of 26~\Ms. 

In some of the progenitors, there will be Si atoms left in the gas after silicate dust is formed to its maximum capacity. Assuming the remaining Si atoms will form silica dust ([SiO$_2$]$_n$), the upper limit of silica mass is compared to the upper limit of silicate dust masses for all progenitors in Figure \ref{fig:Dust_silica}. For this, we did not consider the small Si-rich zone below the O-shell (see Figure \ref{fig:zones_KEPLER}), since most of the pre-explosion Si in that layer is expected to be processed to heavier metals at the time of explosion (as visible in the pre- and post- explosion Si ratio in Figure \ref{fig:zones_KEPLER}). We find that Mg-rich silicate dust dominates the O-rich dust budget until progenitors with an initial mass of 18.5 \Ms. For larger progenitors (initial mass $>$ 18.5 \Ms), the upper limit of silica could reach as large as 0.8 \Ms. Interestingly, very massive progenitors, typically larger than 18 \Ms, are rare for core-collapse SNe (referred to as the red supergiant problem). So the presence/formation of silica dust remains ambiguous. 

Importantly, we find considerable fluctuations in masses of silicate dust in the progenitors with initial masses between 10 and 26 \Ms, as evident in Figure \ref{fig:Dust_KEPLER}. While silicate masses are about 0.1~\Ms\ for stellar progenitors with initial masses of 12--15~\Ms, it is only about 0.025~\Ms\ for progenitors with initial masses between 15--18~\Ms. For progenitors with initial masses above 18 \Ms, the variations are even more dramatic, as we find some clear spikes in silicate mass, which vary between 0.03 to 1.4 \Ms.

%Based on the arguments in Section \ref{sec:dust_conditions}, we do not consider silica as a primary dust type in Type II SNe. 

\begin{figure*}
    \centering

    \begin{subfigure}{0.48\textwidth}
        \includegraphics[width=\linewidth]{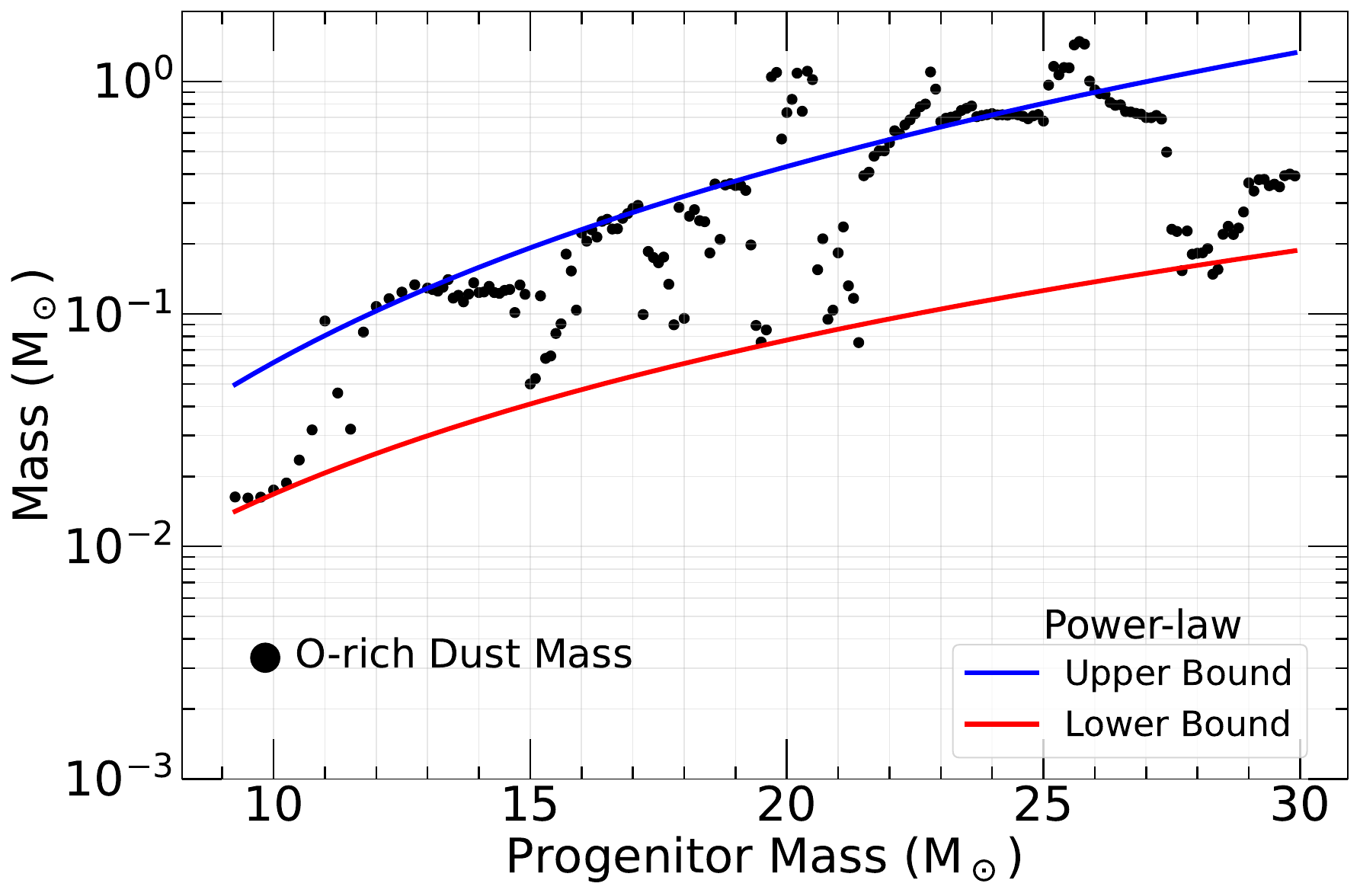}
    \end{subfigure}
    \hfill
    \begin{subfigure}{0.48\textwidth}
         \includegraphics[width=\linewidth]{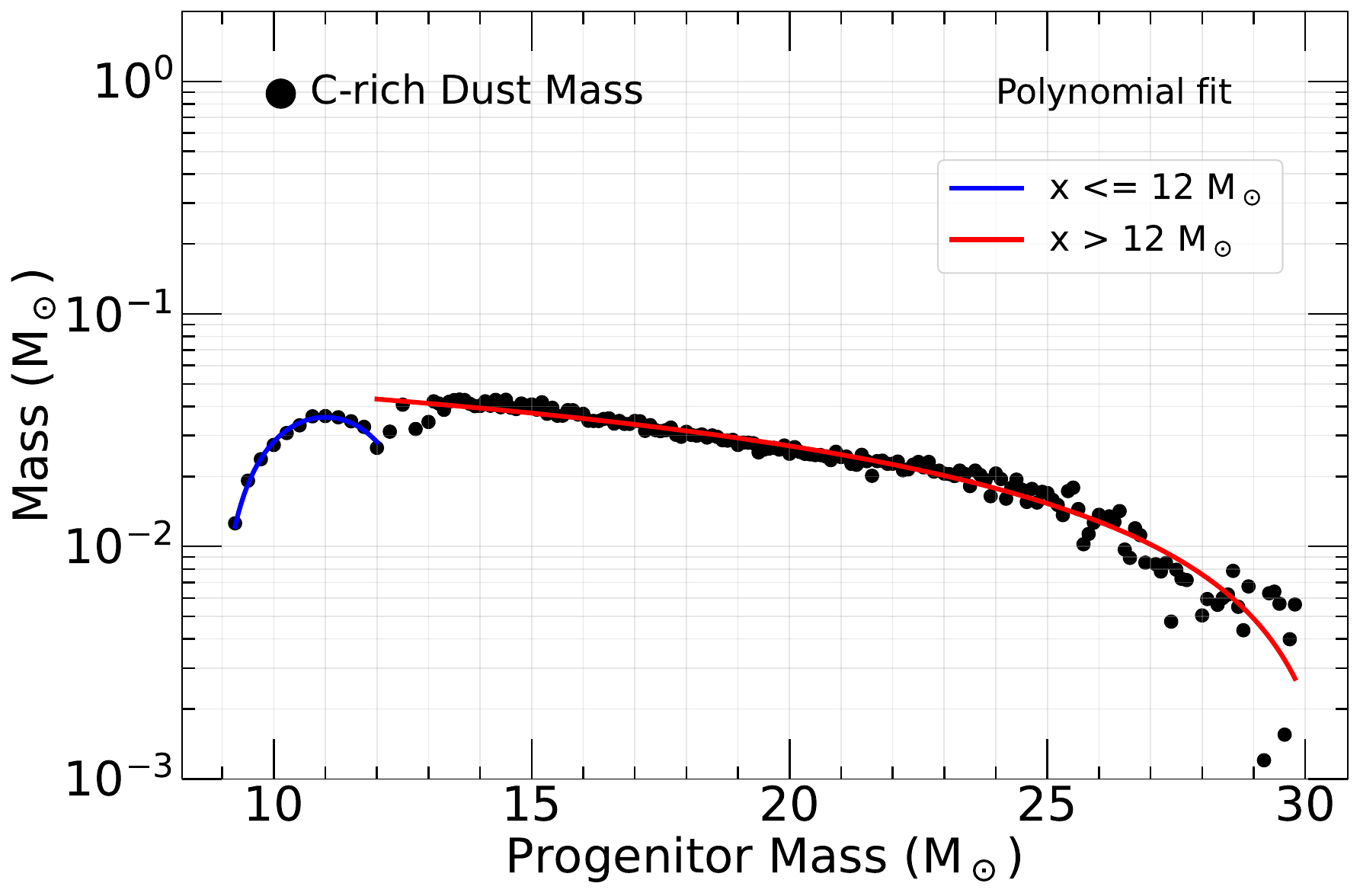}
    \end{subfigure}
    \hfill
    \begin{subfigure}{0.48\textwidth}
         \includegraphics[width=\linewidth]{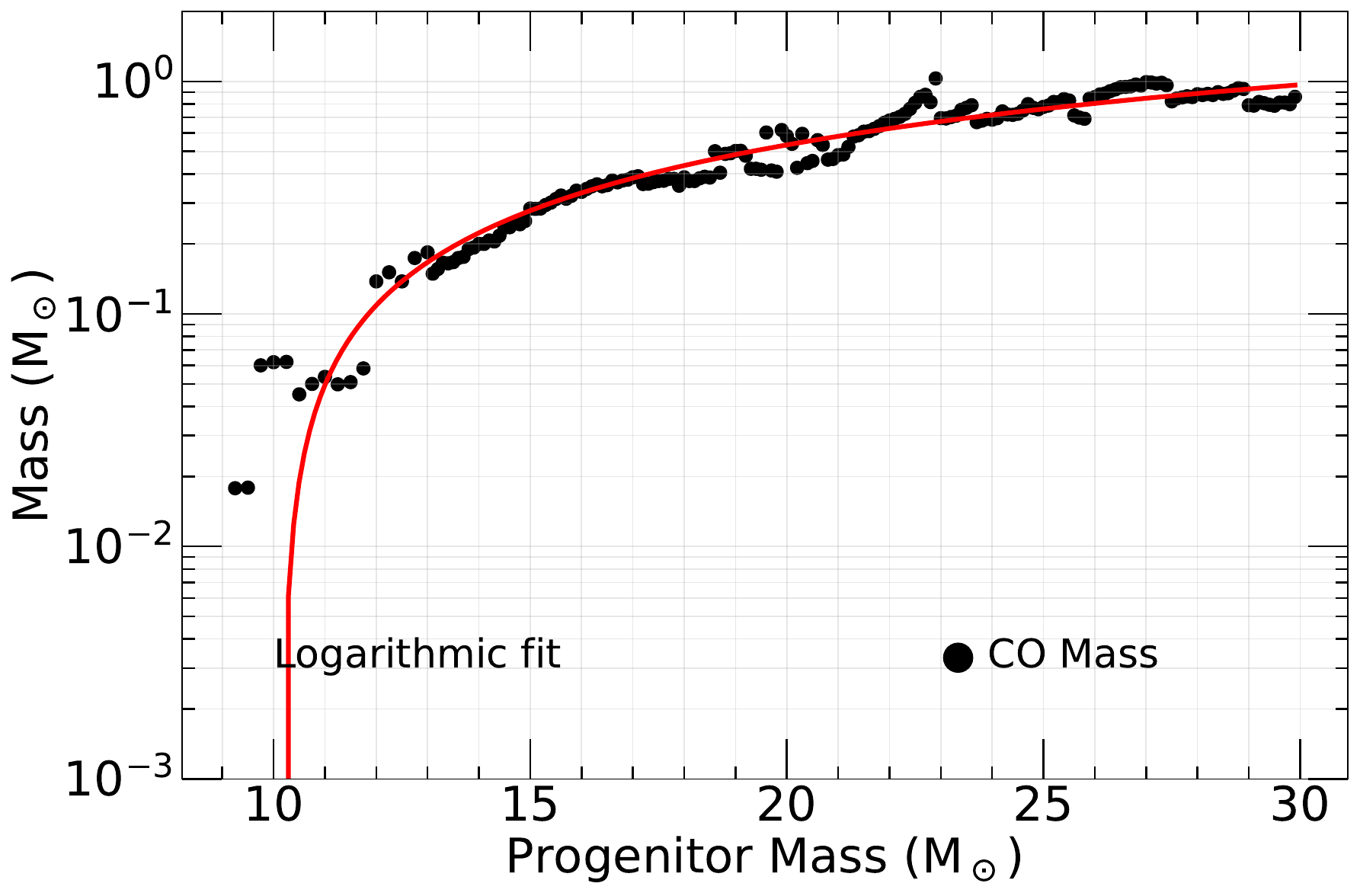}
    \end{subfigure}   
    \hfill
    \begin{subfigure}{0.48\textwidth}
         \includegraphics[width=\linewidth]{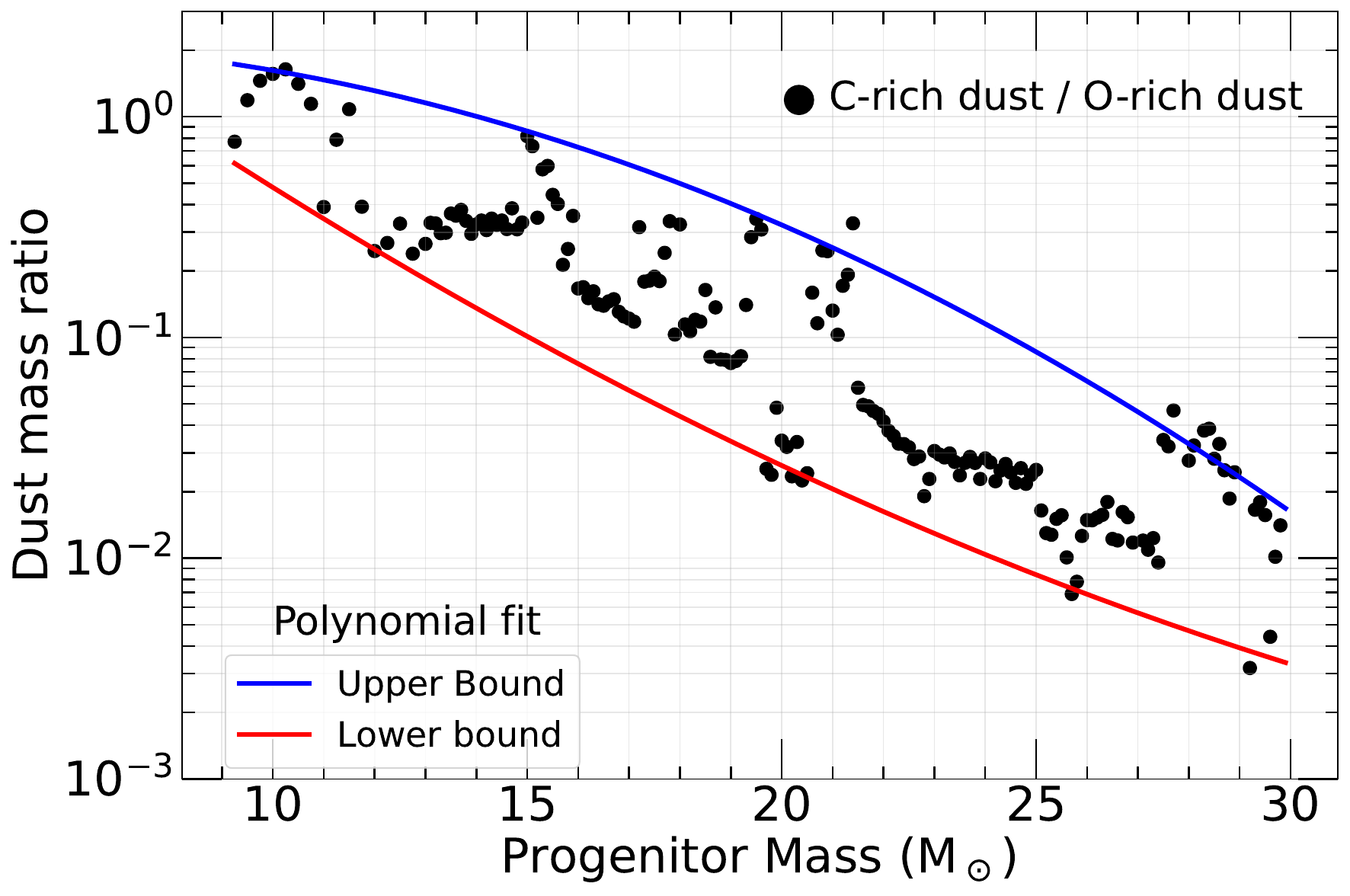}
    \end{subfigure} 
    \caption{The mass of O-rich dust (Top left), C-rich dust (Top right), the ratio of C-rich to O-rich dust (Bottom right), and the mass of CO molecules (Bottom left) across all progenitor masses up to 30~\Ms\ are fitted with various fitting functions. Please see Section \ref{sec_fit_dustmass} for the fitting formulae and constants. } 
     \label{fig:Dust_fit}
\end{figure*}

\subsection{Shell merging and compactness}
\label{sec_compactness}

We find that the large fluctuations in silicate mass can be attributed to the significant fluctuations in the mass of Si and Mg in the O-core, as shown in    Figure~\ref{fig:KEPLER_elemental_mass_compactness}~(top-panel). The effect is especially pronounced in Si mass, which is correlated to the likely convective boundary mixing between the inner Si and O/Ne core, prior to the explosion \citep{davis_2019}. From the yields of \cite{sukhbold_2016}, we find that the shell merging or boundary mixing scenario, as it is termed, leads to an increase in Si abundances in the O/Si/Mg zone to more than an order of magnitude. For example, Figure~\ref{fig:KEPLER_elemental_mass_compactness} (top-panel) shows Si-mass of only 5$\times$10$^{-3}$ \Ms\ in 19.5~\Ms\ progenitor compared to 0.4 \Ms\ in 20.5~\Ms\ progenitor. The degree of shell merging may vary from partial boundary mixing to complete shell mergers. This effect significantly alters the silicate dust masses post-explosion, making dust a crucial tool to probe between different stellar evolutionary channels.

The probability of the shell merging scenario and its markers are extensively studied by \cite{cote_2020, roberti_2025}. There is debate if such stochasticity arises due to numerical effects, rather than being physical. Simulations in 3-D also exhibit similar effects with faster convective velocities \citep{rizzuti_2024}. 

The evolution of the progenitors has also been characterized by a compactness parameter $\xi_{2.5}$, which quantifies how centrally concentrated the mass distribution is within the inner 2.5 M$_\odot$ of a pre-SN star. The compactness parameter is calculated using the mass coordinate where the infall velocity exceeds 1000 km s$^{-1}$ \citep{OConnor2011},

\begin{equation}  \label{eq:compactness}
\xi_{2.5} = \frac{2.5}{R(2.5\,M_\odot)/1000\,\mathrm{km}}
\end{equation}

Using the compactness parameter (shown in Figure~\ref{fig:KEPLER_elemental_mass_compactness}, middle-panel), \citet{Sukhbold2016} demonstrates the likelihood of explosion for the pre-SN models. The less compact pre-SN structures ($\xi_{2.5}$ typically less than 0.28) are assumed to be more likely to explode as SNe \citep{ertl_2016}. 

The derived compactness parameter of a massive star is also correlated to the boundary shell mixing \citep{davis_2019}, where larger degrees of mixing are attributed to smaller values of compactness parameters, therefore more likely to explode as SNe. In this study, we find that the shell mixing leads to an enhancement in silicate dust mass, and therefore, we predict an overall larger mass of dust in those scenarios. These correlations point to the inference that the progenitors, which are more likely to explode as SNe (less compact) will also be capable of producing more dust.   Figure~\ref{fig:KEPLER_elemental_mass_compactness} (middle-panel) shows the comparison of the compactness parameter and O-rich dust mass. A straight forward correlation between the two is not visible in this figure, but can be demonstrated better in the zoomed-in version (19.5 to 21~\Ms) that we present in the bottom-panel of the same figure. The zoomed-in version reflects the anti-correlation of the O-rich dust masses to the compactness of various progenitors, which is then correlated to the shell-merging scenario and explodability.

\subsection{Best-fit dust masses}
\label{sec_fit_dustmass}

To derive a general trend for the dependence of dust mass ($m_d$) on the progenitor's initial mass ($M_s$), we fit the O-rich dust, C-rich dust, and CO molecular mass as a function of progenitor mass. We limit this fitting to progenitors up to 30 \Ms. Here, we provide the best-fit functions, the fitting constants, and their associated errors (expressed as percentages, given in parentheses). Figure \ref{fig:Dust_fit} shows all the best-fit scenarios. 

For O-rich dust, due to the large fluctuations, we could not fit a single function to all the dust masses. The upper and lower bound to the power-law fit is given as, 
\begin{equation} \label{O_powerlaw}
m_d(M_s) = a \times M_s^{b} \quad 
\end{equation}
where $a=9.80\times 10^{-5}, b=2.8$ for upper bound limit and $a=1.06\times 10^{-4}, b=2.2$ for lower bound limit.\\

For C-rich dust, we fit the function in two phases, with a break at $M_s$ = 12 \Ms, as given below, 

\begin{equation}\label{C_polynomial}
m_d(M_s) = -a M_s^{2} + bM_s + c 
\end{equation}
where $a= 7.78\times 10^{-3}(5.33), b=0.171(5.15), c=-0.905(5.15)$ for \( M_s \leq 12 M_\odot \), and
$a= 2.75\times 10^{-5}(26.99), b=-1.11\times 10^{-3}(28.28), c=-0.060(5.32)$ for \( M_s > 12 M_\odot \).\\

The ratio of C-rich to O-rich dust is a crucial indicator of the SN progenitor and explosion properties. Observationally, it is very significant when analyzing IR emission from SN dust. The large scatter in the O-rich dust masses is reflected in the dust mass ratio as well. The polynomial fit is given below:  

\begin{equation} \label{dust_ratio_powerlaw}
log_{10}(m_\mathrm{C/O}(M_s)) = a \times M_s^{2}+b\times M_s+c
\end{equation}

where $a=-3.00\times 10^{-3}, b=0.02, c=0.31$ for upper bound limit and $a=1.80\times 10^{-3}, b=-0.18, c=1.3$ for lower bound limit. \\

For CO molecules, the trend is relatively well defined, with a gradual increase with progenitor mass, given by, 
\begin{equation}\label{CO_logarithmic}
m(M_s) = a \log(M_s - b) + c
\end{equation}
where \ $a= 0.90(42.8), b=-4.50(65.4), c=-2.35(56.1)$.\\

Due to lack of late time observations, CO mass for SNe in general, is not well quantified. Masses of 0.02--1 \Ms\ of CO molecules are reported in SN~1987A \citep{matsuura2017}. In other cases (eg. SN~1998S, SN~2004et, SN~2017eaw, SN~2016adj, SN~2020oi), 10$^{-4}$--10$^{-3}$ \Ms\ of CO were detected in the first couple of years post-explosion, which may not reflect the final mass \citep{rho_2018, kot09, banerjee_2018, rho_2021, fas01}.

\section{Comparison to other stellar yields}
\label{sec_MESA}

The large scatter in the predicted upper limit of dust masses in SNe motivated us to test the sensitivity of our findings using simulation results of the stellar evolution code \texttt{MESA} \citep{paxton_2011, paxton_2015}. 
We apply a similar formalism to the pre-explosion yields of \cite{Laplace2021}, published in Zenodo \citep{laplace_2021_dataset}, which uses code \texttt{MESA}. From this dataset, we derive dust masses for SNe that originate from non-rotating, single star, solar metallicity progenitors with initial masses between 10 and 22 \Ms.

\begin{figure*}
    \centering
    \begin{subfigure}{0.48\textwidth}
        \includegraphics[width=\linewidth]{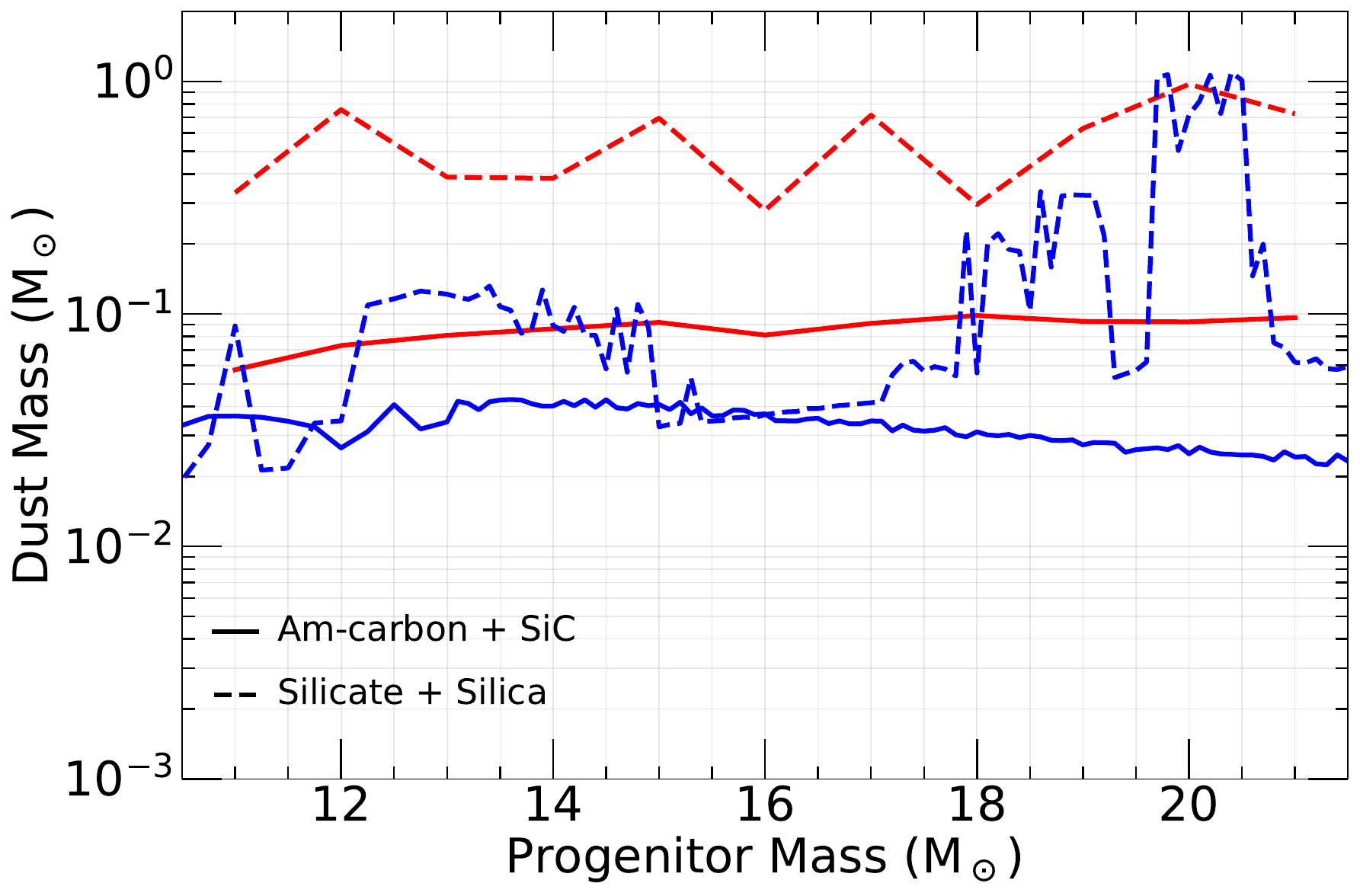}
%        \label{fig:Dust_comparision}
    \end{subfigure}
    \hfill
    \begin{subfigure}{0.48\textwidth}
         \includegraphics[width=\linewidth]{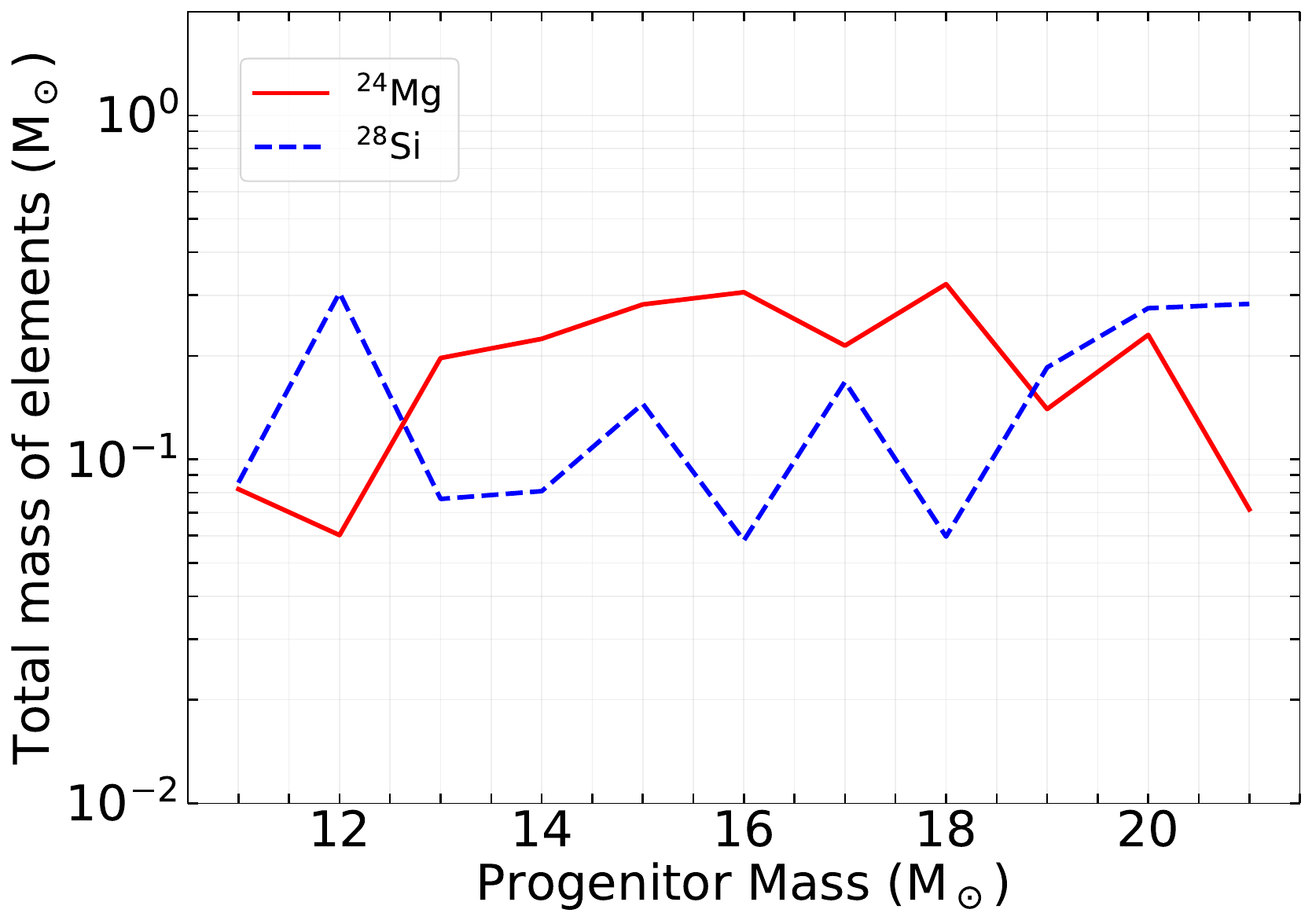}
%        \label{fig:O_shell_MESA}
    \end{subfigure}
    \caption{Left: Comparison between the C-rich (solid lines) and O-rich (dotted lines) dust masses from yields of \citealp{Laplace2021} (code \texttt{MESA}, in red) and \citealp{sukhbold_2016} (code \texttt{KEPLER}, in blue). The larger dust masses in the \citealp{Laplace2021} models are attributed to the larger abundance in the pre-SN data, compared to the \citealp{sukhbold_2016} models. \quad Right: Total mass of elements in the O/Si/Mg zone across progenitor masses from \citealp{Laplace2021}. }
    \label{fig:MESA}
\end{figure*}

We find that the O-rich dust masses still showcase notable fluctuations, with masses between 0.2 and 0.8 \Ms, as shown in Figure \ref{fig:MESA}. Moreover, until the initial mass of 19.5 \Ms, the mass of silicate+silica dust is about 2--5 times larger than what we find using  \citet{sukhbold_2016} models (shown in Figure \ref{fig:Dust_KEPLER}). In this case, O-rich dust always dominates C-rich dust. The final dust masses (both O-rich and C-rich) are larger than the ones based on simulations by \cite{sukhbold_2016}, until the initial mass of 19.5 \Ms. We further compare the Si and Mg abundances, shown in Figure \ref{fig:MESA} (right), to draw parallels with the \cite{sukhbold_2016}  models. In the \cite{Laplace2021} models, Si mass in the O/Si/Mg zone does not vary as widely as in the case of the \cite{sukhbold_2016} models; however, there is still considerable variation that directly reflects in the estimated silicate dust masses. The sample size of initial masses in this dataset \citep{laplace_2021_dataset} is much smaller than \citet{sukhbold_2016}, so the stochasticity of the dust masses is not as pronounced. 

Our finding concludes that dust masses and compositions vary by 2--5 times for the same progenitor mass, based on the stellar model used; it is indeed a huge uncertainty that needs better understanding and finer resolution in initial progenitor masses. 

%We also applied the same methodology to models from \texttt{MESA} simulation and estimated the dust masses across progenitor masses. The O-rich dust masses still contain the sharp features (figure \ref{fig:Dust_KEPLER}). The elevated dust masses in the \texttt{MESA} models are due to the difference in the treatment of the convection and nucleosynthesis network compared to the \texttt{KEPLER} models. From the figure \ref{fig:O_shell_MESA}a, it is clear that the spikes in the O-rich dust masses are not numerical. The Magnesium and Silicon masses in the Oxygen-zone in this case also do not exhibit a clear trend across progenitor masses (figure \ref{fig:O_shell_MESA}a). The irregularities in the mass abundances of Mg and Si (figure  \ref{fig:O_shell_MESA}b) are affecting the O-rich dust formation and is responsible for the spikes in the O-rich in the figure, similar to the estimation from the \texttt{KEPLER} models. The larger dust masses are arising from the larger elemental masses compared to \texttt{KEPLER}.

\section{Conclusions and Discussions}
\label{sec_discussion}

In this paper, we have derived the theoretical upper limit of dust masses that can form in SNe. We find that the final dust masses range between 0.025~\Ms\ to 1.43~\Ms, across progenitors. O-rich dust, especially Mg-silicates, is most often the primary dust component. The general trend suggests that the mass of silicate dust is proportional to the progenitor mass up to about 30~\Ms. The mass of amorphous carbon dust is at its maximum in the smaller progenitors, with initial masses less than 15~\Ms.

\begin{figure*}
    \centering
        \includegraphics[width=16cm]{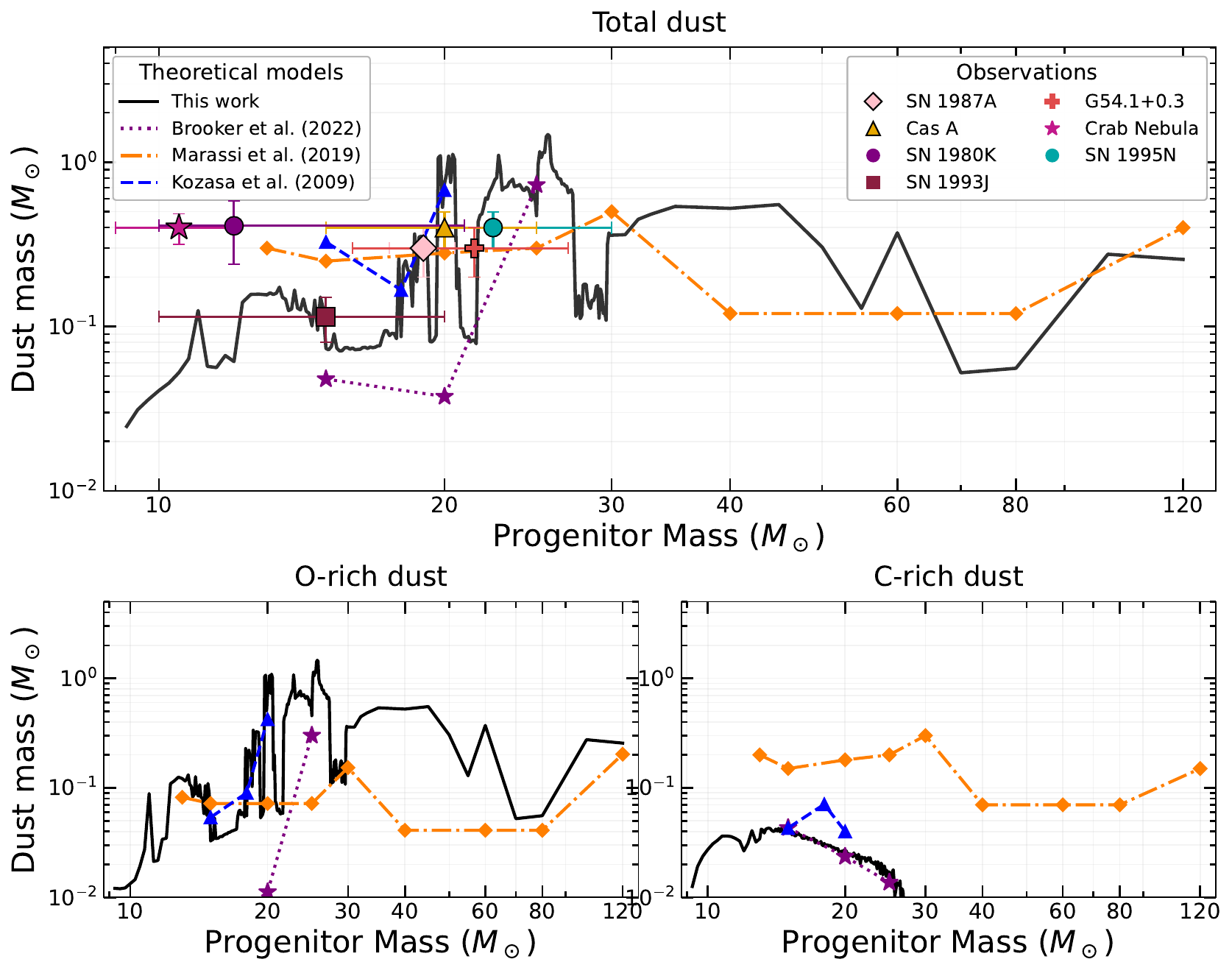}
    
    \caption{Comparison of our results with the dust yields reported by models of \cite{marassi_2019, brooker_2022, kozasa_2009}, for total dust mass (top), mass of O-rich dust (bottom-left), and mass of C-rich dust (bottom-right), as a function of initial stellar mass. We also show the reported dust masses of 30+ yr old SN/SNRs such as SN~1987A \citep{cigan_2019}, Crab Nebula \citep{delooze_2019}, SN~1995N \citep{clayton_2025}, Cas A \citep{delooze2017}, SN~1980K \citep{szanna_2024}, SN~1993J \citep{szalai_2025}, and G54.1+0.3 \citep{temim_2017}. The progenitor mass of observed SN/SNRs are largely uncertain, so the x-axis values of the observed SNe should not be considered as exact.}
    \label{fig:model_compare}
\end{figure*}

Most IR observations of SNe are limited to only the first couple of years post-explosion. It remains uncertain if the observed dust is newly formed or surviving pre-SN dust in the ambient medium. The majority of SNe in those epochs are reported to form dust masses less than 0.01 \Ms (the complete library can be found at \url{https://nebulousresearch.org/dustmasses/}). Mid-IR observations using \textit{JWST} have increased the estimated dust masses, with most SNe found to host dust between 0.01-0.1 \Ms\ \citep{shahbandeh_2023, shahbandeh_2024, szalai_2025, sarangi_2025a}. The presence of colder dust, outside of the mid-IR detection limits, cannot be ruled out. It is generally proposed that all SNe will form close to 0.5 \Ms\ of dust after a few decades post-explosion, based on the detections of cold dust in SNe such as SN~1987A, SN~1995N, Cas A, and SNR G54.1+0.3 \citep{owen_2015, matsuura2017, temim_2017, delooze2017, wesson_2023, clayton_2025}. We have compared the dust masses of several 30+-year-old SNe and SNRs with our model yields in Figure \ref{fig:model_compare}. For these sufficiently old SNe and SNRs (where dust masses should have reached their maximum potential), we find that most of the dust masses compare well with our estimates of the 18--22 \Ms\ initial mass stars. There is significant uncertainty in the progenitor mass of these SNe, so from observations their accurate initial masses cannot be determined.

We reflect on the large scatter in dust masses across progenitors of varying initial mass. We find that the progenitors that underwent convective boundary shell mixing between distinct nuclear burning zones, are associated with larger dust mass production. The progenitors which has undergone shell mixing or merging are likely to be less compact (smaller compactness parameter, \citealp{davis_2019}), and therefore more likely to explode as SNe, and also these are the progenitors that will produce more dust. This can be a reason why many SN remnants are reported to host dust masses larger than 0.5 \Ms, since the stars that exploded as SNe are the ones that are capable of producing large dust masses, post-explosion. However, this explodability factor is mostly important for progenitors above an initial mass of 19~\Ms, below which most of the progenitors are likely to explode as CCSNe \citep{sukhbold_2016}. On the other hand, the shell merging scenario impacts dust masses throughout all progenitor masses, leading to stochasticity in dust mass predictions.

It is also possible that all SNe will not produce dust masses of 0.5 \Ms, or more. From observations with \textit{JWST} in the mid-IR, even at late times, when the ejecta is expected to be optically thin at these wavelengths, we have confirmation of dust masses up to 0.1 \Ms\ \citep{shahbandeh_2023, shahbandeh_2024, sarangi_2025a}. The upper limits derived in this study align better with a 0.1--0.3~\Ms\ range. 

We also show that the upper limits of dust masses, and their compositions, can be very sensitive to the choice of the stellar evolution models, and can vary by 2--5 times for the same progenitor mass. We compared yields from the stellar evolution models of \citealp{sukhbold_2016} (code \texttt{KEPLER}) and \citealp{Laplace2021} (code \texttt{MESA}), which have differences in nuclear reaction network, reaction rates, and energy transport mechanisms \citep{sukhbold_2014, jones_2015}.
The processes that lead to such a random nature of stellar yields and their distributions in the stellar core are poorly understood, which directly reflects on the predicted dust masses. Since most of these alpha elements get locked in dust, we can use observations of dust in the IR as an ideal tool to quantify the stellar yields in the post-explosion era, and thereby constrain the stellar evolutionary channels.

Our results are based on specific 1-D models of non-rotating, solar metallicity, single stars \citep{sukhbold_2016, Laplace2021}. The choice of the stellar evolution model \citep{wang_2024, limongi_2025, sukhbold_2018, eggenberger_2008, rau02, nomoto_1997} or a change in the initial conditions of the star can result in significant variations in predicted dust masses.

In this study, we assumed stratified zones of Si/S, O/Si/Mg, O/C, He/C and H, without microscopic mixing between the layers. From qualitative analysis, we can comment that if we allow mixing between the zones, the mass of C-rich dust is likely to decrease considerably. The free O-atoms in the O/Si/Mg zone will form CO molecules through reactions with carbon atoms in the He/C zone, which in the unmixed case forms amorphous carbon dust. Presence of amorphous carbon dust is often reported from observations of SN dust continuum \citep{wesson_2023, shahbandeh_2023, sarangi_2025a}; so we can argue against microscopic mixing between the zones. The mass of silicate dust is unlikely to increase if the outer zones of Si/O/Mg and He/C are mixed microscopically, since in both the zones, the atoms of Si and Mg are already locked in dust grains in the unmixed case. In that regard, the upper limit of dust masses predicted in this study will remain valid, even if we consider microscopic mixing between zones, post-explosion. 

Theoretical model by \cite{sluder2018} finds MgO-clusters to be the most abundant dust component in SN, dominating over silicates, which boosts the dust mass to 0.5 \Ms. We argue that MgO molecules are not favored by chemistry \citep{bell_2025, bhatt_2007, gobrecht_2023, sar13} due to large energy barriers in the formation reactions. Moreover, signatures of such dust composition are not observationally supported. So we limit our O-rich dust to silicates and alumina.

We have limited our initial stellar models to non-rotating stars at solar metallicity; both these factors are known to impact the abundance structure and dust budget of each source \citep{marassi_2019, fryer_2004, noz03, griffiths_2025, otaki_2025}. We focus on the stochasticity of dust yields derived when a large sample of SN progenitors are considered. Such a large range of initial masses has not been considered in previous liternatures. In Figure \ref{fig:model_compare}, we compare the dust masses derived in this study to the non-rotating solar metallicity models of \cite{marassi_2019} as well as the dust masses predicted by \cite{brooker_2022, kozasa_2009}. The relative fractions of O-rich dust and C-rich dust are also compared for various progenitor masses. We stress that dust formation in SN ejecta is governed by non-equilibrium kinetics, where a large network of thermal and non-thermal processes combine to form dust in a bottom-up formalism \citep{sar13, sarangi2018book}. The dust mass upper-limits derived in this study are tied to this approach. Dust formation in classical nucleation theory relies on the phase transitions, supersaturation ratio, and the partial pressure of the gas phase elements \citep{otaki_2025, brooker_2022, Kozasa2009, marassi_2019, noz10}. However this approach is not applicable in case of rapidly evolving environments like SN ejecta \citep{don85, cherchneff2009}. Moreover, the physical conditions in the SN ejecta are governed by nonthermal processes originating from the radioactive decay of \Ni\ and \Co, which are not accounted for in steady state assumptions. 

In \cite{marassi_2019}, uniform mixing was assumed for the ejecta. We propose mixing only at macroscopic scales, as found in the isotopic signatures of presolar grains and also in the nebular phase spectral analysis \citep{dessart_2020b, hoppe_2024, temim_2019, ergon_2022}. The pre-SN abundances used in \cite{marassi_2019} originate from \cite{limongi_2012, limongi_2017}, where the mass loss for massive stars larger than 30 \Ms\ (initial mass) is treated differently from that of \cite{sukhbold_2016}. That leads to a enhanced C/O ratio in these more massive SN progenitors, reflecting in the C-dust masses found by \cite{marassi_2019} (see Figure \ref{fig:model_compare}). According to \cite{sukhbold_2016}, for such massive progenitors, almost the entire of the He/C shell is lost due to pre-SN mass loss. This leads to a negligible C-dust mass for these progenitors, in our study. \cite{marassi_2019} explores various post-explosion fallback fractions, which alter the production of O-rich dust in the inner ejecta layers. In this work, we focus on the upper-limit of dust masses, so post-explosion fallback was not assumed. \cite{brooker_2022} assumes a stratified ejecta and variation in explosion energy, however the impact of radioactive \Ni\ and \Co\ on the ejecta chemistry was ignored. The impact of radioactivity, or uniform mixing, would alter the abundance of important molecules such as CO, which regulate the formation of C-rich dust.

Our results show that the largest dust masses are expected from stars with initial masses between 20 and 30 \Ms. This finding is in agreement with the models of \cite{marassi_2019, brooker_2022, kozasa_2009}.  

Our results suggest that even though the SN explosion properties, such as explosion energy, \Ni\ mass, clumpiness, or pre-explosion mass-loss rates, affect the timescale of dust formation, but the final mass is controlled by the random nature of the stellar yields. Recent \texttt{JWST} observations of several less-than-decade old CCSNe found no particular correlation between the dust masses and their light curve properties \citep{subrayan_2026}. This may indicate, similar to our finding, that the dust masses are strongly controlled by the stochasticity of the stellar yields, and weakly related by the explosion properties. We need more studies to understand if such stochastic yields of massive stars, as predicted by stellar evolution models, are represented in post-explosion abundances.

\section{Acknowledgments}
The authors gratefully acknowledge the support of the Department of Science and Technology (DST), Government of India. We also extend our sincere thanks to Prof. Stanford Woosley and Prof. Projjwal Banerjee for their valuable feedback and discussions.

%\Jeena{I have added the BibTex entry for Woosley and Weaver (1995), please cite this paper when you mention \textsc{kepler} if you want, as it is one of the relevant papers}

%\Jeena{We use \textsc{kepler} instead of \texttt{KEPLER}. But it is up to you, I just wanted to mention it.}

%\Jeena{How did you get the threshold value 0.2 for the compactness parameter? (is it from the fitting or did Sukhbold mention about it, I will read his paper more carefully) In any case, the shell-merger models have less compactness compared to non-merger models, but it is not mentioned correctly in Sec 7.2, line 431 -- 437}

%\textcolor{blue}{\subsection*{Weighted dust mass}}
\bibliography{library,Bibliography_sarangi_v2}{}

@article{szanna_2024,
	adsurl = {https://ui.adsabs.harvard.edu/abs/2024MNRAS.529..155Z},
	archiveprefix = {arXiv},
	author = {{Zs{\'\i}ros}, Szanna and {Szalai}, Tam{\'a}s and {De Looze}, Ilse and {Sarangi}, Arkaprabha and {Shahbandeh}, Melissa and {Fox}, Ori D. and {Temim}, Tea and {Milisavljevic}, Dan and {Van Dyk}, Schuyler D. and {Smith}, Nathan and {Filippenko}, Alexei V. and {Brink}, Thomas G. and {Zheng}, WeiKang and {Dessart}, Luc and {Jencson}, Jacob and {Johansson}, Joel and {Pierel}, Justin and {Rest}, Armin and {Tinyanont}, Samaporn and {Niculescu-Duvaz}, Maria and {Barlow}, M.~J. and {Wesson}, Roger and {Andrews}, Jennifer and {Clayton}, Geoff and {De}, Kishalay and {Dwek}, Eli and {Engesser}, Michael and {Foley}, Ryan J. and {Gezari}, Suvi and {Gomez}, Sebastian and {Gonzaga}, Shireen and {Kasliwal}, Mansi and {Lau}, Ryan and {Marston}, Anthony and {O'Steen}, Richard and {Siebert}, Matthew and {Skrutskie}, Michael and {Strolger}, Lou and {Wang}, Qinan and {Williams}, Brian and {Williams}, Robert and {Xiao}, Lin},
	doi = {10.1093/mnras/stae507},
	eprint = {2310.03448},
	journal = {\mnras},
	month = mar,
	number = {1},
	pages = {155-168},
	primaryclass = {astro-ph.HE},
	title = {{Serendipitous detection of the dusty Type IIL SN 1980K with JWST/MIRI}},
	volume = {529},
	year = 2024}

@article{niu_2023,
	adsurl = {https://ui.adsabs.harvard.edu/abs/2023ApJ...955L..15N},
	archiveprefix = {arXiv},
	author = {{Niu}, Zexi and {Sun}, Ning-Chen and {Maund}, Justyn R. and {Zhang}, Yu and {Zhao}, Ruining and {Liu}, Jifeng},
	doi = {10.3847/2041-8213/acf4e3},
	eid = {L15},
	eprint = {2308.04677},
	journal = {\apjl},
	month = sep,
	number = {1},
	pages = {L15},
	primaryclass = {astro-ph.SR},
	title = {{The Dusty Red Supergiant Progenitor and the Local Environment of the Type II SN 2023ixf in M101}},
	volume = {955},
	year = 2023}

@article{liu_2023,
	adsurl = {https://ui.adsabs.harvard.edu/abs/2023ApJ...958L..37L},
	archiveprefix = {arXiv},
	author = {{Liu}, Chenxu and {Chen}, Xinlei and {Er}, Xinzhong and {Zeimann}, Gregory R. and {Vink{\'o}}, J{\'o}zsef and {Wheeler}, J. Craig and {Cooper}, Erin Mentuch and {Davis}, Dustin and {Farrow}, Daniel J. and {Gebhardt}, Karl and {Guo}, Helong and {Hill}, Gary J. and {House}, Lindsay and {Kollatschny}, Wolfram and {Kong}, Fanchuan and {Kumar}, Brajesh and {Liu}, Xiangkun and {Tuttle}, Sarah and {Endl}, Michael and {Duke}, Parker and {Cochran}, William D. and {Zhang}, Jinghua and {Liu}, Xiaowei},
	doi = {10.3847/2041-8213/ad0da8},
	eid = {L37},
	eprint = {2311.10400},
	journal = {\apjl},
	month = dec,
	number = {2},
	pages = {L37},
	primaryclass = {astro-ph.GA},
	title = {{The Preexplosion Environments and the Progenitor of SN 2023ixf from the Hobby-Eberly Telescope Dark Energy Experiment (HETDEX)}},
	volume = {958},
	year = 2023}

@article{pledger_2023,
	adsurl = {https://ui.adsabs.harvard.edu/abs/2023ApJ...953L..14P},
	archiveprefix = {arXiv},
	author = {{Pledger}, Joanne L. and {Shara}, Michael M.},
	doi = {10.3847/2041-8213/ace88b},
	eid = {L14},
	eprint = {2305.14447},
	journal = {\apjl},
	month = aug,
	number = {1},
	pages = {L14},
	primaryclass = {astro-ph.SR},
	title = {{Possible Detection of the Progenitor of the Type II Supernova SN 2023ixf}},
	volume = {953},
	year = 2023}

@article{kilpatrick_2023,
	adsurl = {https://ui.adsabs.harvard.edu/abs/2023ApJ...952L..23K},
	archiveprefix = {arXiv},
	author = {{Kilpatrick}, Charles D. and {Foley}, Ryan J. and {Jacobson-Gal{\'a}n}, Wynn V. and {Piro}, Anthony L. and {Smartt}, Stephen J. and {Drout}, Maria R. and {Gagliano}, Alexander and {Gall}, Christa and {Hjorth}, Jens and {Jones}, David O. and {Mandel}, Kaisey S. and {Margutti}, Raffaella and {Ramirez-Ruiz}, Enrico and {Ransome}, Conor L. and {Villar}, V. Ashley and {Coulter}, David A. and {Gao}, Hua and {Matthews}, David Jacob and {Taggart}, Kirsty and {Zenati}, Yossef},
	doi = {10.3847/2041-8213/ace4ca},
	eid = {L23},
	eprint = {2306.04722},
	journal = {\apjl},
	month = jul,
	number = {1},
	pages = {L23},
	primaryclass = {astro-ph.SR},
	title = {{SN 2023ixf in Messier 101: A Variable Red Supergiant as the Progenitor Candidate to a Type II Supernova}},
	volume = {952},
	year = 2023}

@article{jencson_2023,
	adsurl = {https://ui.adsabs.harvard.edu/abs/2023ApJ...952L..30J},
	archiveprefix = {arXiv},
	author = {{Jencson}, Jacob E. and {Pearson}, Jeniveve and {Beasor}, Emma R. and {Lau}, Ryan M. and {Andrews}, Jennifer E. and {Bostroem}, K. Azalee and {Dong}, Yize and {Engesser}, Michael and {Gomez}, Sebastian and {Guolo}, Muryel and {Hoang}, Emily and {Hosseinzadeh}, Griffin and {Jha}, Saurabh W. and {Karambelkar}, Viraj and {Kasliwal}, Mansi M. and {Lundquist}, Michael and {Meza Retamal}, Nicolas E. and {Rest}, Armin and {Sand}, David J. and {Shahbandeh}, Melissa and {Shrestha}, Manisha and {Smith}, Nathan and {Strader}, Jay and {Valenti}, Stefano and {Wang}, Qinan and {Zenati}, Yossef},
	doi = {10.3847/2041-8213/ace618},
	eid = {L30},
	eprint = {2306.08678},
	journal = {\apjl},
	month = aug,
	number = {2},
	pages = {L30},
	primaryclass = {astro-ph.SR},
	title = {{A Luminous Red Supergiant and Dusty Long-period Variable Progenitor for SN 2023ixf}},
	volume = {952},
	year = 2023}

@article{neustadt_2024,
	adsurl = {https://ui.adsabs.harvard.edu/abs/2024MNRAS.527.5366N},
	archiveprefix = {arXiv},
	author = {{Neustadt}, J.~M.~M. and {Kochanek}, C.~S. and {Smith}, M. Rizzo},
	doi = {10.1093/mnras/stad3073},
	eprint = {2306.06162},
	journal = {\mnras},
	month = jan,
	number = {3},
	pages = {5366-5373},
	primaryclass = {astro-ph.HE},
	title = {{Constraints on pre-SN outbursts from the progenitor of SN 2023ixf using the large binocular telescope}},
	volume = {527},
	year = 2024}

@article{xiang_2024,
	adsurl = {https://ui.adsabs.harvard.edu/abs/2024SCPMA..6719514X},
	archiveprefix = {arXiv},
	author = {{Xiang}, Danfeng and {Mo}, Jun and {Wang}, Lingzhi and {Wang}, Xiaofeng and {Zhang}, Jujia and {Lin}, Han and {Wang}, Lifan},
	doi = {10.1007/s11433-023-2267-0},
	eid = {219514},
	eprint = {2309.01389},
	journal = {Science China Physics, Mechanics, and Astronomy},
	month = jan,
	number = {1},
	pages = {219514},
	primaryclass = {astro-ph.SR},
	title = {{The dusty and extremely red progenitor of the type II supernova 2023ixf in Messier 101}},
	volume = {67},
	year = 2024}

@article{vandyk_2024,
	adsurl = {https://ui.adsabs.harvard.edu/abs/2024ApJ...968...27V},
	archiveprefix = {arXiv},
	author = {{Van Dyk}, Schuyler D. and {Srinivasan}, Sundar and {Andrews}, Jennifer E. and {Soraisam}, Monika and {Szalai}, Tam{\'a}s and {Howell}, Steve B. and {Isaacson}, Howard and {Matheson}, Thomas and {Petigura}, Erik and {Scicluna}, Peter and {Stephens}, Andrew W. and {Van Zandt}, Judah and {Zheng}, WeiKang and {Chun}, Sang-Hyun and {Fillippenko}, Alexei V.},
	doi = {10.3847/1538-4357/ad414b},
	eid = {27},
	eprint = {2308.14844},
	journal = {\apj},
	month = jun,
	number = {1},
	pages = {27},
	primaryclass = {astro-ph.SR},
	title = {{The SN 2023ixf Progenitor in M101. II. Properties}},
	volume = {968},
	year = 2024}

@inproceedings{ransome_2024,
	adsurl = {https://ui.adsabs.harvard.edu/abs/2024AAS...24321313R},
	author = {{Ransome}, Conor and {Villar}, V. Ashley and {Jacobson-Galan}, Wynn and {Kilpatrick}, Charles and {Tartaglia}, Anna},
	booktitle = {American Astronomical Society Meeting Abstracts},
	eid = {213.13},
	month = feb,
	pages = {213.13},
	series = {American Astronomical Society Meeting Abstracts},
	title = {{The Pan-STARRS view of the twilight years of the nearby type II supernova, SN2023ixf in the Pinwheel galaxy.}},
	volume = {243},
	year = 2024}

@article{qin_2024,
	adsurl = {https://ui.adsabs.harvard.edu/abs/2024MNRAS.534..271Q},
	archiveprefix = {arXiv},
	author = {{Qin}, Yu-Jing and {Zhang}, Keming and {Bloom}, Joshua and {Sollerman}, Jesper and {Zimmerman}, Erez A. and {Irani}, Ido and {Schulze}, Steve and {Gal-Yam}, Avishay and {Kasliwal}, Mansi and {Coughlin}, Michael W. and {Perley}, Daniel A. and {Fremling}, Christoffer and {Kulkarni}, Shrinivas},
	doi = {10.1093/mnras/stae2012},
	eprint = {2309.10022},
	journal = {\mnras},
	month = oct,
	number = {1},
	pages = {271-280},
	primaryclass = {astro-ph.SR},
	title = {{The progenitor star of SN 2023ixf: a massive red supergiant with enhanced, episodic pre-supernova mass loss}},
	volume = {534},
	year = 2024}

@article{soraisam_2023,
	adsurl = {https://ui.adsabs.harvard.edu/abs/2023ATel16050....1S},
	author = {{Soraisam}, Monika and {Matheson}, Tom and {Andrews}, Jen and {Narayan}, Gautham and {Aleo}, Patrick and {ANTARES Team}},
	journal = {The Astronomer's Telegram},
	month = may,
	pages = {1},
	title = {{Detection of candidate progenitor of SN 2023ixf in HST archival data}},
	volume = {16050},
	year = 2023}

@article{hsu_2024,
	adsurl = {https://ui.adsabs.harvard.edu/abs/2024arXiv240807874H},
	archiveprefix = {arXiv},
	author = {{Hsu}, Brian and {Smith}, Nathan and {Goldberg}, Jared A. and {Bostroem}, K. Azalee and {Hosseinzadeh}, Griffin and {Sand}, David J. and {Pearson}, Jeniveve and {Hiramatsu}, Daichi and {Andrews}, Jennifer E. and {Beasor}, Emma R. and {Dong}, Yize and {Farah}, Joseph and {Galbany}, Llu{\'I}s and {Gomez}, Sebastian and {Padilla Gonzalez}, Estefania and {Guti{\'e}rrez}, Claudia P. and {Howell}, D. Andrew and {K{\"o}nyves-T{\'o}th}, R{\'e}ka and {McCully}, Curtis and {Newsome}, Megan and {Shrestha}, Manisha and {Terreran}, Giacomo and {Villar}, V. Ashley and {Wang}, Xiaofeng},
	doi = {10.48550/arXiv.2408.07874},
	eid = {arXiv:2408.07874},
	eprint = {2408.07874},
	journal = {arXiv e-prints},
	month = aug,
	pages = {arXiv:2408.07874},
	primaryclass = {astro-ph.HE},
	title = {{One Year of SN 2023ixf: Breaking Through the Degenerate Parameter Space in Light-Curve Models with Pulsating Progenitors}},
	year = 2024}

@article{bersten_2024,
	adsurl = {https://ui.adsabs.harvard.edu/abs/2024A&A...681L..18B},
	archiveprefix = {arXiv},
	author = {{Bersten}, M.~C. and {Orellana}, M. and {Folatelli}, G. and {Martinez}, L. and {Piccirilli}, M.~P. and {Regna}, T. and {Rom{\'a}n Aguilar}, L.~M. and {Ertini}, K.},
	doi = {10.1051/0004-6361/202348183},
	eid = {L18},
	eprint = {2310.14407},
	journal = {\aap},
	month = jan,
	pages = {L18},
	primaryclass = {astro-ph.SR},
	title = {{The progenitor of SN 2023ixf from hydrodynamical modeling}},
	volume = {681},
	year = 2024}

@article{moriya_2024,
	adsurl = {https://ui.adsabs.harvard.edu/abs/2024PASJ...76.1050M},
	archiveprefix = {arXiv},
	author = {{Moriya}, Takashi J. and {Singh}, Avinash},
	doi = {10.1093/pasj/psae070},
	eprint = {2406.00928},
	journal = {\pasj},
	month = oct,
	number = {5},
	pages = {1050-1058},
	primaryclass = {astro-ph.HE},
	title = {{Progenitor and explosion properties of SN 2023ixf estimated based on a light-curve model grid of Type II supernovae}},
	volume = {76},
	year = 2024}

@article{singh_2024,
	adsurl = {https://ui.adsabs.harvard.edu/abs/2024ApJ...975..132S},
	archiveprefix = {arXiv},
	author = {{Singh}, Avinash and {Teja}, Rishabh Singh and {Moriya}, Takashi J. and {Maeda}, Keiichi and {Kawabata}, Koji S. and {Tanaka}, Masaomi and {Imazawa}, Ryo and {Nakaoka}, Tatsuya and {Gangopadhyay}, Anjasha and {Yamanaka}, Masayuki and {Swain}, Vishwajeet and {Sahu}, D.~K. and {Anupama}, G.~C. and {Kumar}, Brajesh and {Anche}, Ramya M. and {Sano}, Yasuo and {Raj}, A. and {Agnihotri}, V.~K. and {Bhalerao}, Varun and {Bisht}, D. and {Bisht}, M.~S. and {Belwal}, K. and {Chakrabarti}, S.~K. and {Fujii}, Mitsugu and {Nagayama}, Takahiro and {Matsumoto}, Katsura and {Hamada}, Taisei and {Kawabata}, Miho and {Kumar}, Amit and {Kumar}, Ravi and {Malkan}, Brian K. and {Smith}, Paul and {Sakagami}, Yuta and {Taguchi}, Kenta and {Tominaga}, Nozomu and {Watanabe}, Arata},
	doi = {10.3847/1538-4357/ad7955},
	eid = {132},
	eprint = {2405.20989},
	journal = {\apj},
	month = nov,
	number = {1},
	pages = {132},
	primaryclass = {astro-ph.HE},
	title = {{Unravelling the Asphericities in the Explosion and Multifaceted Circumstellar Matter of SN 2023ixf}},
	volume = {975},
	year = 2024}

@article{ferrari_2024,
	adsurl = {https://ui.adsabs.harvard.edu/abs/2024A&A...687L..20F},
	archiveprefix = {arXiv},
	author = {{Ferrari}, Luc{\'\i}a and {Folatelli}, Gast{\'o}n and {Ertini}, Keila and {Kuncarayakti}, Hanindyo and {Andrews}, Jennifer E.},
	doi = {10.1051/0004-6361/202450440},
	eid = {L20},
	eprint = {2406.00130},
	journal = {\aap},
	month = jul,
	pages = {L20},
	primaryclass = {astro-ph.SR},
	title = {{Progenitor mass and ejecta asymmetry of supernova 2023ixf from nebular spectroscopy}},
	volume = {687},
	year = 2024}

@article{sarangi_2025a,
	adsurl = {https://ui.adsabs.harvard.edu/abs/2025ApJ...993...94S},
	archiveprefix = {arXiv},
	author = {{Sarangi}, Arkaprabha and {Zs{\'\i}ros}, Szanna and {Szalai}, Tam{\'a}s and {Martinez}, Laureano and {Shahbandeh}, Melissa and {Fox}, Ori D. and {Van Dyk}, Schuyler D. and {Filippenko}, Alexei V. and {Bersten}, Melina Cecilia and {De Looze}, Ilse and {Ashall}, Chris and {Temim}, Tea and {Jencson}, Jacob E. and {Rest}, Armin and {Milisavljevic}, Dan and {Dessart}, Luc and {Dwek}, Eli and {Smith}, Nathan and {Tinyanont}, Samaporn and {Brink}, Thomas G. and {Zheng}, WeiKang and {Clayton}, Geoffrey C. and {Andrews}, Jennifer},
	doi = {10.3847/1538-4357/ae0645},
	eid = {94},
	eprint = {2504.20574},
	journal = {\apj},
	month = nov,
	number = {1},
	pages = {94},
	primaryclass = {astro-ph.SR},
	title = {{Two Decades of Dust Evolution in SN 2005af through JWST, Spitzer, and Chemical Modeling}},
	volume = {993},
	year = 2025}

@article{eggenberger_2008,
	adsurl = {https://ui.adsabs.harvard.edu/abs/2008Ap&SS.316...43E},
	author = {{Eggenberger}, P. and {Meynet}, G. and {Maeder}, A. and {Hirschi}, R. and {Charbonnel}, C. and {Talon}, S. and {Ekstr{\"o}m}, S.},
	doi = {10.1007/s10509-007-9511-y},
	journal = {\apss},
	month = aug,
	number = {1-4},
	pages = {43-54},
	title = {{The Geneva stellar evolution code}},
	volume = {316},
	year = 2008}

@ARTICLE{limongi_2012,
       author = {{Limongi}, M. and {Chieffi}, A.},
        title = "{Presupernova Evolution and Explosive Nucleosynthesis of Zero Metal Massive Stars}",
      journal = {\apjs},
         year = 2012,
        month = apr,
       volume = {199},
       number = {2},
          eid = {38},
        pages = {38},
          doi = {10.1088/0067-0049/199/2/38},
archivePrefix = {arXiv},
       eprint = {1202.4581},
 primaryClass = {astro-ph.SR},
       adsurl = {https://ui.adsabs.harvard.edu/abs/2012ApJS..199...38L}
}

@INCOLLECTION{limongi_2017,
       author = {{Limongi}, Marco},
        title = "{Supernovae from Massive Stars}",
    booktitle = {Handbook of Supernovae},
         year = 2017,
       editor = {{Alsabti}, Athem W. and {Murdin}, Paul},
        pages = {513},
          doi = {10.1007/978-3-319-21846-5_119},
       adsurl = {https://ui.adsabs.harvard.edu/abs/2017hsn..book..513L}
}

@article{limongi_2025,
	adsurl = {https://ui.adsabs.harvard.edu/abs/2025ApJS..279...41L},
	archiveprefix = {arXiv},
	author = {{Limongi}, Marco and {Roberti}, Lorenzo and {Falla}, Agnese and {Chieffi}, Alessandro and {Nomoto}, Ken'ichi},
	doi = {10.3847/1538-4365/addf34},
	eid = {41},
	eprint = {2505.22030},
	journal = {\apjs},
	month = aug,
	number = {2},
	pages = {41},
	primaryclass = {astro-ph.SR},
	title = {{The Chemical Yields of Stars in the Mass Range 9{\textendash}15 M$_{{\ensuremath{\odot}}}$}},
	volume = {279},
	year = 2025}

@article{wang_2024,
	adsurl = {https://ui.adsabs.harvard.edu/abs/2024ApJ...962...71W},
	archiveprefix = {arXiv},
	author = {{Wang}, Tianshu and {Burrows}, Adam},
	doi = {10.3847/1538-4357/ad12b8},
	eid = {71},
	eprint = {2311.03446},
	journal = {\apj},
	month = feb,
	number = {1},
	pages = {71},
	primaryclass = {astro-ph.HE},
	title = {{Nucleosynthetic Analysis of Three-dimensional Core-collapse Supernova Simulations}},
	volume = {962},
	year = 2024}

@article{paxton_2011,
	adsurl = {https://ui.adsabs.harvard.edu/abs/2011ApJS..192....3P},
	archiveprefix = {arXiv},
	author = {{Paxton}, B. and {Bildsten}, L. and {Dotter}, A. and {Herwig}, F. and {Lesaffre}, P. and {Timmes}, F.},
	doi = {10.1088/0067-0049/192/1/3},
	eid = {3},
	eprint = {1009.1622},
	journal = {\apjs},
	month = {jan},
	pages = {3},
	primaryclass = {astro-ph.SR},
	title = {{Modules for Experiments in Stellar Astrophysics (MESA)}},
	volume = {192},
	year = {2011}}

@article{paxton_2013,
	adsurl = {https://ui.adsabs.harvard.edu/abs/2013ApJS..208....4P},
	archiveprefix = {arXiv},
	author = {{Paxton}, B. and {Cantiello}, M. and {Arras}, P. and {Bildsten}, L. and {Brown}, E.~F. and {Dotter}, A. and {Mankovich}, C. and {Montgomery}, M.~H. and {Stello}, D. and {Timmes}, F.~X. and {Townsend}, R.},
	doi = {10.1088/0067-0049/208/1/4},
	eid = {4},
	eprint = {1301.0319},
	journal = {\apjs},
	month = {sep},
	pages = {4},
	primaryclass = {astro-ph.SR},
	title = {{Modules for Experiments in Stellar Astrophysics (MESA): Planets, Oscillations, Rotation, and Massive Stars}},
	volume = {208},
	year = {2013}}

@article{paxton_2015,
	adsurl = {https://ui.adsabs.harvard.edu/abs/2015ApJS..220...15P},
	archiveprefix = {arXiv},
	author = {{Paxton}, B. and {Marchant}, P. and {Schwab}, J. and {Bauer}, E.~B. and {Bildsten}, L. and {Cantiello}, M. and {Dessart}, L. and {Farmer}, R. and {Hu}, H. and {Langer}, N. and {Townsend}, R.~H.~D. and {Townsley}, D.~M. and {Timmes}, F.~X.},
	doi = {10.1088/0067-0049/220/1/15},
	eid = {15},
	eprint = {1506.03146},
	journal = {\apjs},
	month = {sep},
	pages = {15},
	primaryclass = {astro-ph.SR},
	title = {{Modules for Experiments in Stellar Astrophysics (MESA): Binaries, Pulsations, and Explosions}},
	volume = {220},
	year = {2015}}

@article{Heger_2003,
	author = {Heger, A. and Fryer, C. L. and Woosley, S. E. and Langer, N. and Hartmann, D. H.},
	doi = {10.1086/375341},
	journal = {\apj},
	month = {jul},
	number = {1},
	pages = {288},
	title = {How Massive Single Stars End Their Life},
	url = {https://dx.doi.org/10.1086/375341},
	volume = {591},
	year = {2003}}

@article{brooker_2022,
	adsurl = {https://ui.adsabs.harvard.edu/abs/2022ApJ...931...85B},
	archiveprefix = {arXiv},
	author = {{Brooker}, Ezra S. and {Stangl}, Sarah M. and {Mauney}, Christopher M. and {Fryer}, C.~L.},
	doi = {10.3847/1538-4357/ac57c3},
	eid = {85},
	eprint = {2103.12781},
	journal = {\apj},
	month = jun,
	number = {2},
	pages = {85},
	primaryclass = {astro-ph.HE},
	title = {{Dependence of Dust Formation on the Supernova Explosion}},
	volume = {931},
	year = 2022}

@article{fryer_2004,
	adsurl = {https://ui.adsabs.harvard.edu/abs/2004ApJ...601..391F},
	archiveprefix = {arXiv},
	author = {{Fryer}, Chris L. and {Warren}, Michael S.},
	doi = {10.1086/380193},
	eprint = {astro-ph/0309539},
	journal = {\apj},
	month = jan,
	number = {1},
	pages = {391-404},
	primaryclass = {astro-ph},
	title = {{The Collapse of Rotating Massive Stars in Three Dimensions}},
	volume = {601},
	year = 2004}

@article{2003A&A...404..975M,
	adsurl = {https://ui.adsabs.harvard.edu/abs/2003A&A...404..975M},
	archiveprefix = {arXiv},
	author = {{Meynet}, G. and {Maeder}, A.},
	doi = {10.1051/0004-6361:20030512},
	eprint = {astro-ph/0304069},
	journal = {\aap},
	month = jun,
	pages = {975-990},
	primaryclass = {astro-ph},
	title = {{Stellar evolution with rotation. X. Wolf-Rayet star populations at solar metallicity}},
	volume = {404},
	year = 2003}

@article{rho_2018,
	adsurl = {https://ui.adsabs.harvard.edu/abs/2018MNRAS.479.5101R},
	archiveprefix = {arXiv},
	author = {{Rho}, J. and {Gomez}, H.~L. and {Boogert}, A. and {Smith}, M.~W.~L. and {Lagage}, P.-O. and {Dowell}, D. and {Clark}, C.~J.~R. and {Peeters}, E. and {Cami}, J.},
	doi = {10.1093/mnras/sty1713},
	eprint = {1707.08230},
	journal = {\mnras},
	month = oct,
	number = {4},
	pages = {5101-5123},
	primaryclass = {astro-ph.GA},
	title = {{A dust twin of Cas A: cool dust and 21 {\ensuremath{\mu}}m silicate dust feature in the supernova remnant G54.1+0.3}},
	volume = {479},
	year = 2018}

@article{rho_2008,
	adsurl = {https://ui.adsabs.harvard.edu/abs/2008ApJ...673..271R},
	archiveprefix = {arXiv},
	author = {{Rho}, J. and {Kozasa}, T. and {Reach}, W.~T. and {Smith}, J.~D. and {Rudnick}, L. and {DeLaney}, T. and {Ennis}, J.~A. and {Gomez}, H. and {Tappe}, A.},
	doi = {10.1086/523835},
	eprint = {0709.2880},
	journal = {\apj},
	month = jan,
	number = {1},
	pages = {271-282},
	primaryclass = {astro-ph},
	title = {{Freshly Formed Dust in the Cassiopeia A Supernova Remnant as Revealed by the Spitzer Space Telescope}},
	volume = {673},
	year = 2008}

@article{otaki_2025,
	adsurl = {https://ui.adsabs.harvard.edu/abs/2025arXiv250611931O},
	archiveprefix = {arXiv},
	author = {{Otaki}, Koki and {Schneider}, Raffaella and {Graziani}, Luca and {Bonella}, Alessandro and {Marassi}, Stefania and {Limongi}, Marco and {Bianchi}, Simone},
	doi = {10.48550/arXiv.2506.11931},
	eid = {arXiv:2506.11931},
	eprint = {2506.11931},
	journal = {arXiv e-prints},
	month = jun,
	pages = {arXiv:2506.11931},
	primaryclass = {astro-ph.GA},
	title = {{Effective supernova dust yields from rotating and non-rotating stellar progenitors}},
	year = 2025}

@inproceedings{kozasa_2009,
	adsurl = {https://ui.adsabs.harvard.edu/abs/2009ASPC..414...43K},
	archiveprefix = {arXiv},
	author = {{Kozasa}, T. and {Nozawa}, T. and {Tominaga}, N. and {Umeda}, H. and {Maeda}, K. and {Nomoto}, K.},
	booktitle = {Cosmic Dust - Near and Far},
	doi = {10.48550/arXiv.0903.0217},
	editor = {{Henning}, T. and {Gr{\"u}n}, E. and {Steinacker}, J.},
	eprint = {0903.0217},
	month = dec,
	pages = {43},
	primaryclass = {astro-ph.GA},
	series = {Astronomical Society of the Pacific Conference Series},
	title = {{Dust in Supernovae: Formation and Evolution}},
	volume = {414},
	year = 2009}

@ARTICLE{cherchneff_2025,
       author = {{Cherchneff}, I. and {Talbi}, D. and {Cernicharo}, J.},
        title = "{Revisiting the formation of molecules and dust in core collapse supernovae}",
      journal = {\aap},
         year = 2026,
        month = mar,
       volume = {708},
          eid = {A76},
        pages = {A76},
          doi = {10.1051/0004-6361/202557490},
archivePrefix = {arXiv},
       eprint = {2510.01079},
 primaryClass = {astro-ph.GA},
       adsurl = {https://ui.adsabs.harvard.edu/abs/2026A&A...708A..76C}
}

@article{bhatt_2007,
	adsurl = {https://ui.adsabs.harvard.edu/abs/2007MNRAS.382..291B},
	author = {{Bhatt}, Jayesh S. and {Ford}, Ian J.},
	doi = {10.1111/j.1365-2966.2007.12358.x},
	journal = {\mnras},
	month = nov,
	number = {1},
	pages = {291-298},
	title = {{Investigation of MgO as a candidate for the primary nucleating dust species around M stars}},
	volume = {382},
	year = 2007}

@article{sukhbold_2016,
	adsurl = {https://ui.adsabs.harvard.edu/abs/2016ApJ...821...38S},
	archiveprefix = {arXiv},
	author = {{Sukhbold}, Tuguldur and {Ertl}, T. and {Woosley}, S.~E. and {Brown}, Justin M. and {Janka}, H. -T.},
	doi = {10.3847/0004-637X/821/1/38},
	eid = {38},
	eprint = {1510.04643},
	journal = {\apj},
	month = apr,
	number = {1},
	pages = {38},
	primaryclass = {astro-ph.HE},
	title = {{Core-collapse Supernovae from 9 to 120 Solar Masses Based on Neutrino-powered Explosions}},
	volume = {821},
	year = 2016}

@article{marassi_2019,
	adsurl = {https://ui.adsabs.harvard.edu/abs/2019MNRAS.484.2587M},
	archiveprefix = {arXiv},
	author = {{Marassi}, S. and {Schneider}, R. and {Limongi}, M. and {Chieffi}, A. and {Graziani}, L. and {Bianchi}, S.},
	doi = {10.1093/mnras/sty3323},
	eprint = {1812.00009},
	journal = {\mnras},
	month = apr,
	number = {2},
	pages = {2587-2604},
	primaryclass = {astro-ph.SR},
	title = {{Supernova dust yields: the role of metallicity, rotation, and fallback}},
	volume = {484},
	year = 2019}

@article{griffiths_2025,
	adsurl = {https://ui.adsabs.harvard.edu/abs/2025A&A...693A..93G},
	archiveprefix = {arXiv},
	author = {{Griffiths}, Adam and {Aloy}, Miguel-{\'A}. and {Hirschi}, Raphael and {Reichert}, Moritz and {Obergaulinger}, Martin and {Whitehead}, Emily E. and {Martinet}, Sebastien and {Sciarini}, Luca and {Ekstr{\"o}m}, Sylvia and {Meynet}, Georges},
	doi = {10.1051/0004-6361/202451816},
	eid = {A93},
	eprint = {2408.03368},
	journal = {\aap},
	month = jan,
	pages = {A93},
	primaryclass = {astro-ph.SR},
	title = {{Evolving massive stars to core collapse with GENEC: Extension of equation of state, opacities and effective nuclear network}},
	volume = {693},
	year = 2025}

@article{clayton_2025,
	adsurl = {https://ui.adsabs.harvard.edu/abs/2025ApJ...991..133C},
	archiveprefix = {arXiv},
	author = {{Clayton}, Geoffrey C. and {Wesson}, R. and {Fox}, Ori D. and {Shahbandeh}, Melissa and {Filippenko}, Alexei V. and {Nickson}, Bryony and {Engesser}, Michael and {Van Dyk}, Schuyler D. and {Zheng}, WeiKang and {Brink}, Thomas G. and {Yang}, Yi and {Temim}, Tea and {Smith}, Nathan and {Andrews}, Jennifer and {Ashall}, Chris and {De Looze}, Ilse and {Derkacy}, James M. and {Dessart}, Luc and {Dulude}, Michael and {Dwek}, Eli and {Foley}, Ryan J. and {Gezari}, Suvi and {Gomez}, Sebastian and {Gonzaga}, Shireen and {Indukuri}, Siva and {Jencson}, Jacob and {Johansson}, Joel and {Kasliwal}, Mansi and {Lane}, Zachary G. and {Lau}, Ryan and {Law}, David and {Marston}, Anthony and {Milisavljevic}, Dan and {O'Steen}, Richard and {Pierel}, Justin and {Rest}, Armin and {Sarangi}, Arkaprabha and {Siebert}, Matthew and {Skrutskie}, Michael and {Strolger}, Lou and {Szalai}, Tam{\'a}s and {Tinyanont}, Samaporn and {Wang}, Qinan and {Williams}, Brian and {Xiao}, Lin and {Zs{\'\i}ros}, Szanna},
	doi = {10.3847/1538-4357/adfc72},
	eid = {133},
	eprint = {2505.01574},
	journal = {\apj},
	month = oct,
	number = {2},
	pages = {133},
	primaryclass = {astro-ph.SR},
	title = {{Very Late-time JWST and Keck Spectra of the Oxygen-rich Supernova 1995N}},
	volume = {991},
	year = 2025}

@article{maria_2021,
	author = {Niculescu-Duvaz, Maria and Barlow, M J and Bevan, A and Milisavljevic, D and De{\^A} Looze, I},
	doi = {10.1093/mnras/stab932},
	eprint = {https://academic.oup.com/mnras/article-pdf/504/2/2133/38118559/stab932.pdf},
	issn = {0035-8711},
	journal = {Monthly Notices of the Royal Astronomical Society},
	month = {04},
	number = {2},
	pages = {2133-2145},
	title = {{The dust mass in Cassiopeia A from infrared and optical line flux differences}},
	url = {https://doi.org/10.1093/mnras/stab932},
	volume = {504},
	year = {2021}}

@article{breemen_2011,
	adsurl = {https://ui.adsabs.harvard.edu/abs/2011A&A...526A.152V},
	archiveprefix = {arXiv},
	author = {{van Breemen}, J.~M. and {Min}, M. and {Chiar}, J.~E. and {Waters}, L.~B.~F.~M. and {Kemper}, F. and {Boogert}, A.~C.~A. and {Cami}, J. and {Decin}, L. and {Knez}, C. and {Sloan}, G.~C. and {Tielens}, A.~G.~G.~M.},
	doi = {10.1051/0004-6361/200811142},
	eid = {A152},
	eprint = {1012.1698},
	journal = {\aap},
	month = feb,
	pages = {A152},
	primaryclass = {astro-ph.GA},
	title = {{The 9.7 and 18 {\ensuremath{\mu}}m silicate absorption profiles towards diffuse and molecular cloud lines-of-sight}},
	volume = {526},
	year = 2011}

@article{arendt_2014,
	author = {Richard G. Arendt and Eli Dwek and Gladys Kober and Jeonghee Rho and Una Hwang},
	doi = {10.1088/0004-637x/786/1/55},
	journal = {The Astrophysical Journal},
	month = {apr},
	number = {1},
	pages = {55},
	publisher = {American Astronomical Society},
	title = {{INTERSTELLAR} {AND} {EJECTA} {DUST} {IN} {THE} {CAS} A {SUPERNOVA} {REMNANT}},
	url = {https://doi.org/10.1088%2F0004-637x%2F786%2F1%2F55},
	volume = {786},
	year = 2014}

@article{temim_2017,
	adsurl = {https://ui.adsabs.harvard.edu/abs/2017ApJ...836..129T},
	archiveprefix = {arXiv},
	author = {{Temim}, Tea and {Dwek}, Eli and {Arendt}, Richard G. and {Borkowski}, Kazimierz J. and {Reynolds}, Stephen P. and {Slane}, Patrick and {Gelfand}, Joseph D. and {Raymond}, John C.},
	doi = {10.3847/1538-4357/836/1/129},
	eid = {129},
	eprint = {1701.01117},
	journal = {\apj},
	month = feb,
	number = {1},
	pages = {129},
	primaryclass = {astro-ph.GA},
	title = {{A Massive Shell of Supernova-formed Dust in SNR G54.1+0.3}},
	volume = {836},
	year = 2017}

@article{davis_2019,
	adsurl = {https://ui.adsabs.harvard.edu/abs/2019MNRAS.484.3921D},
	archiveprefix = {arXiv},
	author = {{Davis}, A. and {Jones}, S. and {Herwig}, F.},
	doi = {10.1093/mnras/sty3415},
	eprint = {1712.00114},
	journal = {\mnras},
	month = apr,
	number = {3},
	pages = {3921-3934},
	primaryclass = {astro-ph.SR},
	title = {{Convective boundary mixing in a post-He core burning massive star model}},
	volume = {484},
	year = 2019}

@article{roberti_2025,
	adsurl = {https://ui.adsabs.harvard.edu/abs/2025A&A...698A.216R},
	archiveprefix = {arXiv},
	author = {{Roberti}, L. and {Pignatari}, M. and {Brinkman}, H.~E. and {Jeena}, S.~K. and {Sieverding}, A. and {Falla}, A. and {Limongi}, M. and {Chieffi}, A. and {Lugaro}, M.},
	doi = {10.1051/0004-6361/202554461},
	eid = {A216},
	eprint = {2504.18867},
	journal = {\aap},
	month = jun,
	pages = {A216},
	primaryclass = {astro-ph.SR},
	title = {{The occurrence and impact of carbon-oxygen shell mergers in massive stars}},
	volume = {698},
	year = 2025}

@article{ertl_2016,
	adsurl = {https://ui.adsabs.harvard.edu/abs/2016ApJ...818..124E},
	archiveprefix = {arXiv},
	author = {{Ertl}, T. and {Janka}, H. -Th. and {Woosley}, S.~E. and {Sukhbold}, T. and {Ugliano}, M.},
	doi = {10.3847/0004-637X/818/2/124},
	eid = {124},
	eprint = {1503.07522},
	journal = {\apj},
	month = feb,
	number = {2},
	pages = {124},
	primaryclass = {astro-ph.SR},
	title = {{A Two-parameter Criterion for Classifying the Explodability of Massive Stars by the Neutrino-driven Mechanism}},
	volume = {818},
	year = 2016}

@article{rizzuti_2024,
	adsurl = {https://ui.adsabs.harvard.edu/abs/2024MNRAS.533..687R},
	archiveprefix = {arXiv},
	author = {{Rizzuti}, F. and {Hirschi}, R. and {Varma}, V. and {Arnett}, W.~D. and {Georgy}, C. and {Meakin}, C. and {Moc{\'a}k}, M. and {Murphy}, A. StJ and {Rauscher}, T.},
	doi = {10.1093/mnras/stae1778},
	eprint = {2407.15544},
	journal = {\mnras},
	month = sep,
	number = {1},
	pages = {687-704},
	primaryclass = {astro-ph.SR},
	title = {{Shell mergers in the late stages of massive star evolution: new insight from 3D hydrodynamic simulations}},
	volume = {533},
	year = 2024}

@article{cote_2020,
	adsurl = {https://ui.adsabs.harvard.edu/abs/2020ApJ...892...57C},
	archiveprefix = {arXiv},
	author = {{C{\^o}t{\'e}}, Benoit and {Jones}, Samuel and {Herwig}, Falk and {Pignatari}, Marco},
	doi = {10.3847/1538-4357/ab77ac},
	eid = {57},
	eprint = {1906.07218},
	journal = {\apj},
	month = mar,
	number = {1},
	pages = {57},
	primaryclass = {astro-ph.SR},
	title = {{Chromium Nucleosynthesis and Silicon-Carbon Shell Mergers in Massive Stars}},
	volume = {892},
	year = 2020}

@article{smartt_2009,
	adsurl = {https://ui.adsabs.harvard.edu/abs/2009ARA&A..47...63S},
	archiveprefix = {arXiv},
	author = {{Smartt}, Stephen J.},
	doi = {10.1146/annurev-astro-082708-101737},
	eprint = {0908.0700},
	journal = {\araa},
	month = sep,
	number = {1},
	pages = {63-106},
	primaryclass = {astro-ph.SR},
	title = {{Progenitors of Core-Collapse Supernovae}},
	volume = {47},
	year = 2009}

@article{wesson_2021,
	adsurl = {https://ui.adsabs.harvard.edu/abs/2021ApJ...923..148W},
	archiveprefix = {arXiv},
	author = {{Wesson}, Roger and {Bevan}, Antonia},
	doi = {10.3847/1538-4357/ac2eb8},
	eid = {148},
	eprint = {2111.06896},
	journal = {\apj},
	month = dec,
	number = {2},
	pages = {148},
	primaryclass = {astro-ph.SR},
	title = {{Observational Limits on the Early-time Dust Mass in SN 1987A}},
	volume = {923},
	year = 2021}

@ARTICLE{dessart_2020b,
       author = {{Dessart}, L. and {Hillier}, D. John},
        title = "{Radiative-transfer modeling of supernovae in the nebular-phase. A novel treatment of chemical mixing in spherical symmetry}",
      journal = {\aap},
         year = 2020,
        month = nov,
       volume = {643},
          eid = {L13},
        pages = {L13},
          doi = {10.1051/0004-6361/202039287},
archivePrefix = {arXiv},
       eprint = {2011.02862},
 primaryClass = {astro-ph.HE},
       adsurl = {https://ui.adsabs.harvard.edu/abs/2020A&A...643L..13D}
}

@article{sukhbold_2018,
	adsurl = {https://ui.adsabs.harvard.edu/abs/2018ApJ...860...93S},
	archiveprefix = {arXiv},
	author = {{Sukhbold}, Tuguldur and {Woosley}, S.~E. and {Heger}, Alexander},
	doi = {10.3847/1538-4357/aac2da},
	eid = {93},
	eprint = {1710.03243},
	journal = {\apj},
	month = jun,
	number = {2},
	pages = {93},
	primaryclass = {astro-ph.HE},
	title = {{A High-resolution Study of Presupernova Core Structure}},
	volume = {860},
	year = 2018}

@article{delooze_2019,
	adsurl = {https://ui.adsabs.harvard.edu/abs/2019MNRAS.488..164D},
	archiveprefix = {arXiv},
	author = {{De Looze}, I. and {Barlow}, M.~J. and {Bandiera}, R. and {Bevan}, A. and {Bietenholz}, M.~F. and {Chawner}, H. and {Gomez}, H.~L. and {Matsuura}, M. and {Priestley}, F. and {Wesson}, R.},
	doi = {10.1093/mnras/stz1533},
	eprint = {1906.02203},
	journal = {\mnras},
	month = sep,
	number = {1},
	pages = {164-182},
	primaryclass = {astro-ph.HE},
	title = {{The dust content of the Crab Nebula}},
	volume = {488},
	year = 2019}

@article{dwek_2019,
	adsurl = {https://ui.adsabs.harvard.edu/abs/2019ApJ...871L..33D},
	archiveprefix = {arXiv},
	author = {{Dwek}, Eli and {Sarangi}, Arkaprabha and {Arendt}, Richard G.},
	doi = {10.3847/2041-8213/aaf9a8},
	eid = {L33},
	eprint = {1812.08234},
	journal = {\apjl},
	month = feb,
	number = {2},
	pages = {L33},
	primaryclass = {astro-ph.SR},
	title = {{The Evolution of Dust Opacity in Core Collapse Supernovae and the Rapid Formation of Dust in Their Ejecta}},
	volume = {871},
	year = 2019}

@ARTICLE{ergon_2022,
       author = {{Ergon}, Mattias and {Fransson}, Claes},
        title = "{Spectral modelling of Type IIb supernovae. Comparison with SN 2011dh and the effect of macroscopic mixing}",
      journal = {\aap},
         year = 2022,
        month = oct,
       volume = {666},
          eid = {A104},
        pages = {A104},
          doi = {10.1051/0004-6361/202243448},
archivePrefix = {arXiv},
       eprint = {2206.11014},
 primaryClass = {astro-ph.SR},
       adsurl = {https://ui.adsabs.harvard.edu/abs/2022A&A...666A.104E}
}

@ARTICLE{temim_2019,
       author = {{Temim}, Tea and {Slane}, Patrick and {Sukhbold}, Tuguldur and {Koo}, Bon-Chul and {Raymond}, John C. and {Gelfand}, Joseph D.},
        title = "{Probing the Innermost Ejecta Layers in Supernova Remnant Kes 75: Implications for the Supernova Progenitor}",
      journal = {\apjl},
         year = 2019,
        month = jun,
       volume = {878},
       number = {1},
          eid = {L19},
        pages = {L19},
          doi = {10.3847/2041-8213/ab237c},
archivePrefix = {arXiv},
       eprint = {1905.02849},
 primaryClass = {astro-ph.HE},
       adsurl = {https://ui.adsabs.harvard.edu/abs/2019ApJ...878L..19T}
}

@ARTICLE{hoppe_2024,
       author = {{Hoppe}, Peter and {Leitner}, Jan and {Pignatari}, Marco and {Amari}, Sachiko},
        title = "{Isotope studies of presolar silicon carbide grains from supernovae: new constraints for hydrogen-ingestion supernova models}",
      journal = {\mnras},
         year = 2024,
        month = jul,
       volume = {532},
       number = {1},
        pages = {211-222},
          doi = {10.1093/mnras/stae1523},
       adsurl = {https://ui.adsabs.harvard.edu/abs/2024MNRAS.532..211H}
}

@article{gobrecht_2023,
	adsurl = {https://ui.adsabs.harvard.edu/abs/2023A&A...680A..18G},
	archiveprefix = {arXiv},
	author = {{Gobrecht}, David and {Hashemi}, Seyyed Rasoul and {Plane}, John Maurice Campbell and {Bromley}, Stefan Thomas and {Nyman}, Gunnar and {Decin}, Leen},
	doi = {10.1051/0004-6361/202347546},
	eid = {A18},
	eprint = {2310.08521},
	journal = {\aap},
	month = dec,
	pages = {A18},
	primaryclass = {astro-ph.SR},
	title = {{Bottom-up dust nucleation theory in oxygen-rich evolved stars. II. Magnesium and calcium aluminate clusters}},
	volume = {680},
	year = 2023}

@article{bell_2025,
	address = {Department of Chemistry and Biochemistry, University of Mississippi, University, MS 38677, USA.; Department of Chemistry and Biochemistry, University of Mississippi, University, MS 38677, USA.},
	auid = {ORCID: 0000-0003-4716-8225},
	author = {Bell, Kailey M and Fortenberry, Ryan C},
	cois = {The authors declare no conflicts of interest.},
	crdt = {2025/05/07 13:24},
	date = {2025 Apr 8},
	dep = {20250408},
	doi = {10.3390/molecules30081650},
	edat = {2025/05/07 18:47},
	gr = {NNH2ZHA004C/NASA/NASA/United States},
	issn = {1420-3049 (Electronic); 1420-3049 (Linking)},
	jid = {100964009},
	journal = {Molecules},
	jt = {Molecules (Basel, Switzerland)},
	language = {eng},
	lid = {10.3390/molecules30081650 {$[$}doi{$]$}; 1650},
	lr = {20250509},
	mhda = {2025/05/07 18:48},
	month = {Apr},
	number = {8},
	oto = {NOTNLM},
	own = {NLM},
	phst = {2025/03/10 00:00 {$[$}received{$]$}; 2025/04/01 00:00 {$[$}revised{$]$}; 2025/04/03 00:00 {$[$}accepted{$]$}; 2025/05/07 18:48 {$[$}medline{$]$}; 2025/05/07 18:47 {$[$}pubmed{$]$}; 2025/05/07 13:24 {$[$}entrez{$]$}; 2025/04/08 00:00 {$[$}pmc-release{$]$}},
	pii = {molecules30081650; molecules-30-01650},
	pl = {Switzerland},
	pmc = {PMC12029866},
	pmcr = {2025/04/08},
	pmid = {40333547},
	pst = {epublish},
	pt = {Journal Article},
	sb = {IM},
	status = {PubMed-not-MEDLINE},
	title = {The Formation of MgS \& MgO Monomers and Dimers from Magnesium, Oxygen, and Sulfur Hydrides.},
	volume = {30},
	year = {2025}}

@article{jones_2015,
	adsurl = {https://ui.adsabs.harvard.edu/abs/2015MNRAS.447.3115J},
	archiveprefix = {arXiv},
	author = {{Jones}, S. and {Hirschi}, R. and {Pignatari}, M. and {Heger}, A. and {Georgy}, C. and {Nishimura}, N. and {Fryer}, C. and {Herwig}, F.},
	doi = {10.1093/mnras/stu2657},
	eprint = {1412.6518},
	journal = {\mnras},
	month = mar,
	number = {4},
	pages = {3115-3129},
	primaryclass = {astro-ph.SR},
	title = {{Code dependencies of pre-supernova evolution and nucleosynthesis in massive stars: evolution to the end of core helium burning}},
	volume = {447},
	year = 2015}

@article{laplace_2021_dataset,
	author = {Laplace, E. and Justham, S. and Renzo, M. and G{\"o}tberg, Y. and Farmer, R. and Vartanyan, D. and de Mink, S. E.},
	doi = {10.5281/zenodo.5556959},
	month = feb,
	publisher = {Zenodo},
	title = {Different to the core: the pre-supernova structures of massive single and binary-stripped stars},
	url = {https://doi.org/10.5281/zenodo.5556959},
	version = {2.0.1},
	year = 2021}

@article{sukhbold_2014,
	adsurl = {https://ui.adsabs.harvard.edu/abs/2014ApJ...783...10S},
	archiveprefix = {arXiv},
	author = {{Sukhbold}, Tuguldur and {Woosley}, S.~E.},
	doi = {10.1088/0004-637X/783/1/10},
	eid = {10},
	eprint = {1311.6546},
	journal = {\apj},
	month = mar,
	number = {1},
	pages = {10},
	primaryclass = {astro-ph.SR},
	title = {{The Compactness of Presupernova Stellar Cores}},
	volume = {783},
	year = 2014}

@article{owen_2015,
	adsurl = {https://ui.adsabs.harvard.edu/abs/2015ApJ...801..141O},
	archiveprefix = {arXiv},
	author = {{Owen}, P.~J. and {Barlow}, M.~J.},
	doi = {10.1088/0004-637X/801/2/141},
	eid = {141},
	eprint = {1501.01510},
	journal = {\apj},
	month = mar,
	number = {2},
	pages = {141},
	primaryclass = {astro-ph.GA},
	title = {{The Dust and Gas Content of the Crab Nebula}},
	volume = {801},
	year = 2015}

@article{wesson_2023,
	adsurl = {https://ui.adsabs.harvard.edu/abs/2023MNRAS.525.4928W},
	archiveprefix = {arXiv},
	author = {{Wesson}, R. and {Bevan}, A.~M. and {Barlow}, M.~J. and {De Looze}, I. and {Matsuura}, M. and {Clayton}, G. and {Andrews}, J.},
	doi = {10.1093/mnras/stad2505},
	eprint = {2308.09028},
	journal = {\mnras},
	month = nov,
	number = {4},
	pages = {4928-4941},
	primaryclass = {astro-ph.SR},
	title = {{Evidence for late-time dust formation in the ejecta of supernova SN 1995N from emission-line asymmetries}},
	volume = {525},
	year = 2023}

@article{shahbandeh_2024,
	adsurl = {https://ui.adsabs.harvard.edu/abs/2025ApJ...985..262S},
	archiveprefix = {arXiv},
	author = {{Shahbandeh}, Melissa and {Fox}, Ori D. and {Temim}, Tea and {Dwek}, Eli and {Sarangi}, Arkaprabha and {Smith}, Nathan and {Dessart}, Luc and {Nickson}, Bryony and {Engesser}, Michael and {Filippenko}, Alexei V. and {Brink}, Thomas G. and {Zheng}, WeiKang and {Szalai}, Tam{\'a}s and {Johansson}, Joel and {Rest}, Armin and {Van Dyk}, Schuyler D. and {Andrews}, Jennifer and {Ashall}, Chris and {Clayton}, Geoffrey C. and {De Looze}, Ilse and {DerKacy}, James M. and {Dulude}, Michael and {Foley}, Ryan J. and {Gezari}, Suvi and {Gomez}, Sebastian and {Gonzaga}, Shireen and {Indukuri}, Siva and {Jencson}, Jacob and {Kasliwal}, Mansi and {Lane}, Zachary G. and {Lau}, Ryan and {Law}, David and {Marston}, Anthony and {Milisavljevic}, Dan and {O'Steen}, Richard and {Pierel}, Justin and {Siebert}, Matthew and {Skrutskie}, Michael and {Strolger}, Lou and {Tinyanont}, Samaporn and {Wang}, Qinan and {Williams}, Brian and {Xiao}, Lin and {Yang}, Yi and {Zs{\'\i}ros}, Szanna},
	doi = {10.3847/1538-4357/adce77},
	eid = {262},
	eprint = {2410.09142},
	journal = {\apj},
	month = jun,
	number = {2},
	pages = {262},
	primaryclass = {astro-ph.HE},
	title = {{JWST/MIRI Observations of Newly Formed Dust in the Cold, Dense Shell of the Type IIn SN 2005ip}},
	volume = {985},
	year = 2025}

@article{shahbandeh_2023,
	adsurl = {https://ui.adsabs.harvard.edu/abs/2023MNRAS.523.6048S},
	archiveprefix = {arXiv},
	author = {{Shahbandeh}, Melissa and {Sarangi}, Arkaprabha and {Temim}, Tea and {Szalai}, Tam{\'a}s and {Fox}, Ori D. and {Tinyanont}, Samaporn and {Dwek}, Eli and {Dessart}, Luc and {Filippenko}, Alexei V. and {Brink}, Thomas G. and {Foley}, Ryan J. and {Jencson}, Jacob and {Pierel}, Justin and {Zs{\'\i}ros}, Szanna and {Rest}, Armin and {Zheng}, WeiKang and {Andrews}, Jennifer and {Clayton}, Geoffrey C. and {De}, Kishalay and {Engesser}, Michael and {Gezari}, Suvi and {Gomez}, Sebastian and {Gonzaga}, Shireen and {Johansson}, Joel and {Kasliwal}, Mansi and {Lau}, Ryan and {De Looze}, Ilse and {Marston}, Anthony and {Milisavljevic}, Dan and {O'Steen}, Richard and {Siebert}, Matthew and {Skrutskie}, Michael and {Smith}, Nathan and {Strolger}, Lou and {Van Dyk}, Schuyler D. and {Wang}, Qinan and {Williams}, Brian and {Williams}, Robert and {Xiao}, Lin and {Yang}, Yi},
	doi = {10.1093/mnras/stad1681},
	eprint = {2301.10778},
	journal = {\mnras},
	month = aug,
	number = {4},
	pages = {6048-6060},
	primaryclass = {astro-ph.HE},
	title = {{JWST observations of dust reservoirs in type IIP supernovae 2004et and 2017eaw}},
	volume = {523},
	year = 2023}

@article{szalai_2019,
	author = {Tam{\'{a}}s Szalai and Szanna Zs{\'{\i}}ros and Ori D. Fox and Ondr{\v}ej Pejcha and Tom{\'{a}}s M{\"u}ller},
	doi = {10.3847/1538-4365/ab10df},
	journal = {The Astrophysical Journal Supplement Series},
	month = {apr},
	number = {2},
	pages = {38},
	publisher = {American Astronomical Society},
	title = {A Comprehensive Analysis of Spitzer Supernovae},
	url = {https://doi.org/10.3847%2F1538-4365%2Fab10df},
	volume = {241},
	year = 2019}

@article{sluder2018,
	adsurl = {http://adsabs.harvard.edu/abs/2018MNRAS.480.5580S},
	archiveprefix = {arXiv},
	author = {{Sluder}, A. and {Milosavljevi{\'c}}, M. and {Montgomery}, M.~H.},
	doi = {10.1093/mnras/sty2060},
	eprint = {1612.09013},
	journal = {\mnras},
	month = nov,
	pages = {5580-5624},
	primaryclass = {astro-ph.SR},
	title = {{Molecular nucleation theory of dust formation in core-collapse supernovae applied to SN 1987A}},
	volume = 480,
	year = 2018}

@article{delooze2017,
	adsurl = {http://adsabs.harvard.edu/abs/2017MNRAS.465.3309D},
	archiveprefix = {arXiv},
	author = {{De Looze}, I. and {Barlow}, M.~J. and {Swinyard}, B.~M. and {Rho}, J. and {Gomez}, H.~L. and {Matsuura}, M. and {Wesson}, R.},
	doi = {10.1093/mnras/stw2837},
	eprint = {1611.00774},
	journal = {\mnras},
	month = mar,
	pages = {3309-3342},
	title = {{The dust mass in Cassiopeia A from a spatially resolved Herschel analysis}},
	volume = 465,
	year = 2017}

@article{sarangi2018book,
	adsurl = {http://adsabs.harvard.edu/abs/2018SSRv..214...63S},
	author = {{Sarangi}, A. and {Matsuura}, M. and {Micelotta}, E.~R.},
	doi = {10.1007/s11214-018-0492-7},
	eid = {63},
	journal = {\ssr},
	month = apr,
	pages = {63},
	title = {{Dust in Supernovae and Supernova Remnants I: Formation Scenarios}},
	volume = 214,
	year = 2018}

@inbook{matsuura2017,
	adsurl = {http://adsabs.harvard.edu/abs/2017hsn..book.2125M},
	author = {{Matsuura}, M.},
	booktitle = {Handbook of Supernovae, ISBN 978-3-319-21845-8.~Springer International Publishing AG, 2017, p.~2125},
	doi = {10.1007/978-3-319-21846-5_130},
	pages = {2125},
	publisher = {Springer},
	title = {{Dust and Molecular Formation in Supernovae}},
	year = 2017}

@article{shahbandeh_SN2005ip,
	adsurl = {https://ui.adsabs.harvard.edu/abs/2024arXiv241009142S},
	archiveprefix = {arXiv},
	author = {{Shahbandeh}, Melissa and {Fox}, Ori D. and {Temim}, Tea and {Dwek}, Eli and {Sarangi}, Arkaprabha and {Smith}, Nathan and {Dessart}, Luc and {Nickson}, Bryony and {Engesser}, Michael and {Filippenko}, Alexei V. and {Brink}, Thomas G. and {Zheng}, Weikang and {Szalai}, Tam{\'a}s and {Johansson}, Joel and {Rest}, Armin and {Van Dyk}, Schuyler D. and {Andrews}, Jennifer and {Ashall}, Chris and {Clayton}, Geoffrey C. and {De Looze}, Ilse and {Derkacy}, James M. and {Dulude}, Michael and {Foley}, Ryan J. and {Gezari}, Suvi and {Gomez}, Sebastian and {Gonzaga}, Shireen and {Indukuri}, Siva and {Jencson}, Jacob and {Kasliwal}, Mansi and {Lane}, Zachary G. and {Lau}, Ryan and {Law}, David and {Marston}, Anthony and {Milisavljevic}, Dan and {O'Steen}, Richard and {Pierel}, Justin and {Siebert}, Matthew and {Skrutskie}, Michael and {Strolger}, Lou and {Tinyanont}, Samaporn and {Wang}, Qinan and {Williams}, Brian and {Xiao}, Lin and {Yang}, Yi and {Zs{\'\i}ros}, Szanna},
	doi = {10.48550/arXiv.2410.09142},
	eid = {arXiv:2410.09142},
	eprint = {2410.09142},
	journal = {arXiv e-prints},
	month = oct,
	pages = {arXiv:2410.09142},
	primaryclass = {astro-ph.HE},
	title = {{JWST/MIRI Observations of Newly Formed Dust in the Cold, Dense Shell of the Type IIn SN 2005ip}},
	year = 2024}

@article{rau02,
	adsurl = {http://adsabs.harvard.edu/abs/2002ApJ...576..323R},
	author = {{Rauscher}, T. and {Heger}, A. and {Hoffman}, R.~D. and {Woosley}, S.~E.},
	doi = {10.1086/341728},
	eprint = {astro-ph/0112478},
	journal = {The Astrophysical Journal},
	month = sep,
	pages = {323-348},
	title = {{Nucleosynthesis in Massive Stars with Improved Nuclear and Stellar Physics}},
	volume = 576,
	year = 2002}

@article{cherchneff2009,
	adsurl = {http://adsabs.harvard.edu/abs/2009ApJ...703..642C},
	archiveprefix = {arXiv},
	author = {{Cherchneff}, I. and {Dwek}, E.},
	doi = {10.1088/0004-637X/703/1/642},
	eprint = {0907.3621},
	journal = {The Astrophysical Journal},
	month = sep,
	pages = {642-661},
	primaryclass = {astro-ph.SR},
	title = {{The Chemistry of Population III Supernova Ejecta. I. Formation of Molecules in the Early Universe}},
	volume = 703,
	year = 2009}

@article{sar13,
	adsurl = {http://adsabs.harvard.edu/abs/2013ApJ...776..107S},
	archiveprefix = {arXiv},
	author = {{Sarangi}, A. and {Cherchneff}, I.},
	doi = {10.1088/0004-637X/776/2/107},
	eid = {107},
	eprint = {1309.5887},
	journal = {The Astrophysical Journal},
	month = oct,
	pages = {107},
	primaryclass = {astro-ph.SR},
	title = {{The Chemically Controlled Synthesis of Dust in Type II-P Supernovae}},
	volume = 776,
	year = 2013}

@article{kot09,
	adsurl = {http://adsabs.harvard.edu/abs/2009ApJ...704..306K},
	archiveprefix = {arXiv},
	author = {{Kotak}, R. and {Meikle}, W.~P.~S. and {Farrah}, D. and {Gerardy}, C.~L. and {Foley}, R.~J. and {Van Dyk}, S.~D. and {Fransson}, C. and {Lundqvist}, P. and {Sollerman}, J. and {Fesen}, R. and {Filippenko}, A.~V. and {Mattila}, S. and {Silverman}, J.~M. and {Andersen}, A.~C. and {H{\"o}flich}, P.~A. and {Pozzo}, M. and {Wheeler}, J.~C.},
	doi = {10.1088/0004-637X/704/1/306},
	eprint = {0904.3737},
	journal = {The Astrophysical Journal},
	month = oct,
	pages = {306-323},
	primaryclass = {astro-ph.SR},
	title = {{Dust and The Type II-Plateau Supernova 2004et}},
	volume = 704,
	year = 2009}

@ARTICLE{szalai_2025,
       author = {{Szalai}, Tam{\'a}s and {Zs{\'\i}ros}, Szanna and {Jencson}, Jacob and {Fox}, Ori D. and {Shahbandeh}, Melissa and {Sarangi}, Arkaprabha and {Temim}, Tea and {De Looze}, Ilse and {Smith}, Nathan and {Filippenko}, Alexei V. and {Van Dyk}, Schuyler D. and {Andrews}, Jennifer and {Ashall}, Chris and {Clayton}, Geoffrey C. and {Dessart}, Luc and {Dulude}, Michael and {Dwek}, Eli and {Gomez}, Sebastian and {Johansson}, Joel and {Milisavljevic}, Dan and {Pierel}, Justin and {Rest}, Armin and {Tinyanont}, Samaporn and {Brink}, Thomas G. and {De}, Kishalay and {Engesser}, Michael and {Foley}, Ryan J. and {Gezari}, Suvi and {Kasliwal}, Mansi and {Lau}, Ryan and {Marston}, Anthony and {O'Steen}, Richard and {Siebert}, Matthew and {Skrutskie}, Michael and {Strolger}, Lou and {Wang}, Qinan and {Williams}, Brian J. and {Williams}, Robert and {Xiao}, Lin and {Zheng}, WeiKang},
        title = "{JWST/MIRI detects the dusty SN1993J about 30 years after explosion}",
      journal = {\aap},
         year = 2025,
        month = may,
       volume = {697},
          eid = {A132},
        pages = {A132},
          doi = {10.1051/0004-6361/202451470},
archivePrefix = {arXiv},
       eprint = {2503.12950},
 primaryClass = {astro-ph.SR},
       adsurl = {https://ui.adsabs.harvard.edu/abs/2025A&A...697A.132S}
}

@article{noz10,
	adsurl = {http://adsabs.harvard.edu/abs/2010ApJ...713..356N},
	archiveprefix = {arXiv},
	author = {{Nozawa}, T. and {Kozasa}, T. and {Tominaga}, N. and {Maeda}, K. and {Umeda}, H. and {Nomoto}, K. and {Krause}, O.},
	doi = {10.1088/0004-637X/713/1/356},
	eprint = {0909.4145},
	journal = {The Astrophysical Journal},
	month = apr,
	pages = {356-373},
	primaryclass = {astro-ph.SR},
	title = {{Formation and Evolution of Dust in Type IIb Supernovae with Application to the Cassiopeia A Supernova Remnant}},
	volume = 713,
	year = 2010}

@article{noz03,
	adsurl = {http://adsabs.harvard.edu/abs/2003ApJ...598..785N},
	author = {{Nozawa}, T. and {Kozasa}, T. and {Umeda}, H. and {Maeda}, K. and {Nomoto}, K.},
	doi = {10.1086/379011},
	eprint = {astro-ph/0307108},
	journal = {The Astrophysical Journal},
	month = dec,
	pages = {785-803},
	title = {{Dust in the Early Universe: Dust Formation in the Ejecta of Population III Supernovae}},
	volume = 598,
	year = 2003}

@article{fas01,
	adsurl = {http://adsabs.harvard.edu/abs/2001MNRAS.325..907F},
	author = {{Fassia}, A. and {Meikle}, W.~P.~S. and {Chugai}, N. and {Geballe}, T.~R. and {Lundqvist}, P. and {Walton}, N.~A. and {Pollacco}, D. and {Veilleux}, S. and {Wright}, G.~S. and {Pettini}, M. and {Kerr}, T. and {Puchnarewicz}, E. and {Puxley}, P. and {Irwin}, M. and {Packham}, C. and {Smartt}, S.~J. and {Harmer}, D.},
	doi = {10.1046/j.1365-8711.2001.04282.x},
	eprint = {astro-ph/0011340},
	journal = {Monthly Notices of the Royal Astronomical Society},
	month = aug,
	pages = {907-930},
	title = {{Optical and infrared spectroscopy of the type IIn SN 1998S: days 3-127}},
	volume = 325,
	year = 2001}

@article{weaver_1978,
	adsurl = {https://ui.adsabs.harvard.edu/abs/1978ApJ...225.1021W},
	author = {{Weaver}, T.~A. and {Zimmerman}, G.~B. and {Woosley}, S.~E.},
	doi = {10.1086/156569},
	journal = {\apj},
	month = nov,
	pages = {1021-1029},
	title = {{Presupernova evolution of massive stars.}},
	volume = {225},
	year = 1978}

@article{woo02,
	adsurl = {http://adsabs.harvard.edu/abs/2002RvMP...74.1015W},
	author = {{Woosley}, S.~E. and {Heger}, A. and {Weaver}, T.~A.},
	doi = {10.1103/RevModPhys.74.1015},
	journal = {Reviews of Modern Physics},
	month = nov,
	pages = {1015-1071},
	title = {{The evolution and explosion of massive stars}},
	volume = 74,
	year = 2002}

@article{Kozasa2009,
   author = {Takashi Kozasa and Takaya Nozawa and Nozomu Tominaga and Hideyuki Umeda and Keiichi Maeda and Ken'ichi Nomoto},
   month = {3},
   title = {Dust in Supernovae; Formation and Evolution},
   year = {2009}
}
\bibliographystyle{aasjournalv7}

%% The "ht!" tells LaTeX to put the figure "here" first, at the "top" next
%% and to override the normal way of calculating a float position.
%% The asterisk after "figure" tells the compiler to span multiple columns
%% if a two column style is selected.

%To invoke a two column style similar to what is produced in
%the published PDF copy use: \\

%\noindent {\tt\string\documentclass[twocolumn]\{aastex7\}}. \\

\end{document}